# Submillimeter Galaxies at z~2: Evidence for Major Mergers & Constraints on Lifetimes, IMF and CO-H$_2$ conversion factor[1]

L.J. Tacconi[(1)], R. Genzel[(1,2)], I. Smail[(3)], R. Neri[(4)], S.C. Chapman[(5)], R. J. Ivison[(6)], A. Blain[(7)], P. Cox[(4)], A. Omont[(8)], F. Bertoldi[(9)], T. Greve[(10)], N.M. Förster Schreiber[(1)], S. Genel[(1)], D. Lutz[(1)], A.M. Swinbank[(3)], A.E. Shapley[(11)], D.K. Erb[(12)], A. Cimatti[(13)], E. Daddi[(14)] & A.J. Baker[(15)]

(1) Max-Planck Institut für extraterrestrische Physik, (MPE), Giessenbachstrasse 1, D-85741 Garching, Germany (linda@mpe.mpg.de, genzel@mpe.mpg.de, forster@mpe.mpg.de, shy@mpe.mpg.de, lutz@mpe.mpg.de)
(2) Department of Physics, University of California, Le Conte Hall, Berkeley, CA, 94720 USA
(3) Institute for Computational Cosmology, Durham University, Durham, United Kingdom (Ian.Smail@durham.ac.uk, a.m.swinbank@dur.ac.uk)
(4) Institut de Radio Astronomie Millimetrique (IRAM), St.Martin d'Heres, France (neri@iram.fr, Pierre.Cox@iram.fr)
(5) Institute of Astronomy, University of Cambridge, Madingley Road, Cambridge, CB3 0HA, United Kingdom (schapman@ast.cam.ac.uk)
(6) UK Astronomy Technology Centre, Royal Observatory, Blackford Hill, Edinburgh EH9 3HJ, United Kingdom and Institute for Astronomy, University of Edinburgh, Blackford Hill, Edinburgh EH9 3HJ, United Kingdom (rji@roe.ac.uk)
(7) Astronomy 105-24, California Institute of Technology, Pasadena, CA 91125 USA (awb@astro.caltech.edu, schapman@irastro.caltech.edu, tgreve@submm.caltech.edu)
(8) CNRS & Institut d'Astrophysique de Paris, 98 bis boulevard Arago, 75014 Paris, France (omont@iap.fr)
(9) AIUB, Bonn, Germany (bertoldi@astro.uni-bonn.de)
(10) Max-Planck Institut für Astronomie (MPIA), Königsstuhl 17, D-68117 Heidelberg, Germany (tgreve@mpia.de)
(11) Department of Astrophysical Sciences, Princeton University, Peyton Hall, Princeton, NJ 08544 USA (aes@astro.princeton.edu)
(12) Harvard-Smithsonian Center for Astrophysics, 60 Garden St., Cambridge, MA 02138, USA (derb@cfa.harvard.edu)
(13) Dipartimento di Astronomia – Alma Mater Studiorum – Università di Bologna, Via Ranzani 1, I-40127 Bologna, Italy (a.cimatti@unibo.it)
(14) Laboratoire AIM, CEA/DSM - CNRS – Université Paris Diderot, DAPNIA/SAp, Orme des Merisiers, 91191 Gif-sur-Yvette, France (emanuele.daddi@cea.fr)
(15) Dept. of Physics & Astronomy, Rutgers, the State University of NJ,136 Frelinghuysen Road Piscataway, NJ 08854 USA (ajbaker@physics.rutgers.edu)

## *Abstract*

We report sub-arcsecond resolution IRAM PdBI millimeter CO interferometry of four

z~2 submillimeter galaxies (SMGs), and sensitive CO (3-2) flux limits toward three z~2

UV-/optically selected star forming galaxies. The new data reveal for the first time spatially resolved CO gas kinematics in the observed SMGs. Two of the SMGs show double or multiple morphologies, with complex, disturbed gas motions. The other two SMGs exhibit CO velocity gradients of ~500 km s$^{-1}$ across ≤0.2" (1.6 kpc) diameter regions, suggesting that the star forming gas is in compact, rotating disks. Our data provide compelling evidence that these SMGs represent extreme, short-lived 'maximum' star forming events in highly dissipative mergers of gas rich galaxies. The resulting high mass surface and volume densities of SMGs are similar to those of compact quiescent galaxies in the same redshift range, and much higher than those in local spheroids. From the ratio of the comoving volume densities of SMGs and quiescent galaxies in the same mass and redshift ranges, and from the comparison of gas exhaustion time scales and stellar ages, we estimate that the SMG phase duration is about 100 Myrs. Our analysis of SMGs and optically/UV selected high redshift star forming galaxies supports a 'universal' Chabrier IMF as being valid over the star forming history of these galaxies. We find that the $^{12}$CO luminosity to total gas mass conversion factors at z~2-3 are probably similar to those assumed at z~0. The implied gas fractions in our sample galaxies range from 20 to 50%.

---

[1] Based on observations obtained at the IRAM Plateau de Bure Interferometer (PdBI). IRAM is funded by the Centre National de la Recherché Scientifique (France), the Max-Planck Gesellschaft (Germany), and the Instituto Geografico Nacional (Spain).

# 1. Introduction

Deep surveys have become efficient in detecting and studying z~1.5-3 star forming galaxy populations (e.g. Steidel et al. 1996, 2004, Franx et al. 2003, Daddi et al. 2004b, Hughes et al. 1998, Chapman et al. 2005). Fairly large samples are now available based on their rest-frame UV magnitude/color properties, for example the z~3 Lyman break galaxies (LBGs), (e.g. Steidel et al. 1996), and the z~1.4-2.5 BX/BM galaxies (e.g. Adelberger et al. 2004, Steidel et al. 2004, Erb et al. 2006a). Other samples have been chosen based on their rest-frame optical magnitude/color properties, such as the z~1.4-2.6 'star forming' or 's'-BzKs (e.g. Daddi et al. 2004a,b, Kong et al. 2006), or distant red galaxies (DRGs) (Franx et al. 2003). Finally, dusty luminous high redshift galaxies at z~1-3.5 have been unveiled through their high submillimeter flux densities (e.g. Smail et al. 2002; Chapman et al. 2003,2005; Pope et al. 2005). These bright, submillimeter continuum-selected systems (SMGs) with $S_{850\mu m}>5$mJy are dusty and gas rich, very luminous (~$10^{13}$ $L_\odot$) galaxies with star formation rates, SFR~$10^3$ $M_\odot$yr$^{-1}$ (Blain et al. 2002, Smail et al. 2002, Chapman et al. 2005). In contrast the BX/BM or s-BzK/DRG selection criteria sample luminous (L~$10^{11-12}$ $L_\odot$) galaxies with SFR~10-500 $M_\odot$ yr$^{-1}$, and with estimated ages from 50 Myrs to 2 Gyrs (e.g. Erb et al. 2006a,b,c, Förster Schreiber et al. 2004, Daddi et al. 2004a,b, Papovich et al. 2007).

For the SMGs, Chapman et al. (2003, 2005) and Swinbank et al. (2004) have been able to obtain rest-frame UV/optical redshifts for 73 SMGs detected with the SCUBA and MAMBO cameras at 850μm and 1.2mm. In comparison to earlier attempts, this effort has succeeded thanks to precise positions derived from deep 1.4 GHz VLA observations of the same fields (Ivison et al. 2002; Chapman et al. 2003). Starting from these results some of us have been carrying out a survey at the IRAM Plateau de Bure millimeter interferometer (PdBI), where we successfully detected molecular emission and determined line profiles in the CO (3-2) and (4-3) rotational lines for eighteen of these SMGs (Neri et al. 2003, Greve et al. 2005, Tacconi et al. 2006, Smail et al. in prep.). Tacconi et al. (2006) presented sub-arcsecond resolution PdBI millimeter imaging of six of the program SMGs, showed that several of these sources were surprisingly compact ($R_{1/2} \leq 2$ kpc) and that SMGs are 'maximum starbursts', that is systems in which a significant fraction of the available initial gas reservoir of $10^{10}$–$10^{11}$ $M_\odot$ is converted to stars over several dynamical timescales, typically of the order of a few times $10^8$ yr. The SMGs appear to be scaled-up versions of the local ultra-luminous infrared galaxy population (Sanders & Mirabel 1996).

Significant progress has been made in elucidating the physical properties of high-z star forming galaxies, as well as the relations of different populations to each other (e.g. Reddy et al. 2005, Grazian et al. 2007, Bouché et al. 2007). However, one important open issue is the stellar initial mass function (IMF) at high redshift (e.g. Renzini 2005, van Dokkum 2007, Dave 2007). An empirical determination of the IMF (slope, lower and upper mass range, mass to light ratio) in high-z star forming galaxies would obviously be of substantial interest for future galaxy evolution studies. Various arguments based on the $M_*/L_B$ ratio of $z \sim 0$ ellipticals, the slope and zero point of the $z=0-1.3$ fundamental plane of spheroids in galaxy clusters and the metal abundances in galaxy clusters suggest that the Milky Way IMF (Kroupa 2001, Chabrier 2003) may be universally applicable, at least to moderate redshifts (Renzini 2005). On the other hand, there is good evidence that in the massive, dense Arches and NGC3603 star clusters and the Galactic Center nuclear cluster the IMF is top heavy. For NGC3603 Stolte et al. (2002) and Harayama, Eisenhauer & Martins (2007) find a power law slope of the IMF of $\gamma=1.7^2$, where Salpeter, Kroupa, and Chabrier IMFs have $\gamma=2.3-2.35$ for $m \geq 1$ $M_\odot$. In the central parsec of the Milky Way, Nayakshin & Sunyaev (2005) and Paumard et al. (2006) infer $\gamma\sim1$. Baugh et al. (2005) conclude that a top heavy IMF ($\gamma\sim1$) is required to fit the number counts of SMGs in their semi-analytic models (Lacey et al. 2007). Without a normal Galactic IMF the Baugh et al. (2005) models miss the observed number counts by factors of between 20 and 40. Extending these models to include metallicities and supernova feedback, Nagashima et al. (2005) conclude that a top heavy IMF also better fits the observed super-solar [O/Fe] abundance ratios of massive ellipticals. More recently Lacey et al. (2007) confirm and extend this conclusion when fitting in addition the luminosity function of Spitzer 24-160 μm counts. Finally, van Dokkum (2007) and Dave (2007) discuss further possible evidence and motivation for the IMF to be more top-heavy at high redshift than it is locally. To test these proposals it is obviously of interest to obtain a direct constraint on the high-z IMF.

In the present paper we follow up on the work of Tacconi et al. (2006) and increased the angular resolution of the PdBI observations by another factor of 2, to 0.25"-0.5" FWHM. For four of the z~2 SMGs we now resolve the compact sources *and* their CO kinematics. We find compelling kinematic evidence for highly dissipative mergers and orbital motion on scales of <1kpc and set new constraints on the physical properties and evolution of these SMGs. We also report sensitive limits on CO emission toward three representatives of the UV-/optically star forming population at z~2. For the standard $H=70$ km s$^{-1}$ Mpc$^{-1}$, $\Omega_m=0.3$ $\Lambda$CDM cosmology adopted here 1 arcsecond corresponds to 8.2 kpc at z=2.2.

---

[2] Here we define the IMF in terms of the number of stars in the mass interval between $m_*$, and $m_*+dm_*$. For a single power law IMF this number would be: $dN(m_*)\sim m_*^{-\gamma} dm_*$

## 2. Observations

The submillimeter galaxy observations were carried out in two seasons (winter 2005/2006 and 2006/2007) with the PdBI (Guilloteau et al. 1992), which consists of six 15m-diameter telescopes. For our high resolution observations in the new extended A configuration (760 meter baseline) we selected four galaxies of the $S_{850\mu m} \geq 5$ mJy radio detected SCUBA sample (Chapman et al. 2005). We had previously observed all four galaxies with the PdBI in the shorter baseline configurations (Neri et al. (2003), Greve et al. (2005), Tacconi et al. (2006)). Weather conditions during the observations were excellent. We observed the galaxies HDF242 (SMMJ123707+6214, z=2.49, Chapman et al. 2003, 2005, Swinbank et al. 2004, Tacconi et al. 2006) and N2850.4 (SMMJ16350+4057, z=2.39, Ivison et al. 2002, Smail et al. 2003, Neri et al. 2003, Chapman et al. 2005) in January-February 2006. For this period the array was equipped with both (single- sideband) 3 mm and (double-sideband) 1 mm SIS receivers that we used simultaneously. System temperatures (referred to above the atmosphere) were 110-160 K and 350-500 K at 3 and 1 mm, respectively. We observed HDF76 (SMMJ123549+6215, z=2.20, Chapman et al. 2005, Swinbank et al. 2004, Tacconi et al. 2006) and N2850.2 (SMMJ163658+4105, z=2.45, Ivison et al. 2002, Greve et al. 2005, Chapman et al. 2005, Tacconi et al. 2006) in February 2007 with IRAM's new dual-polarization receivers at 1mm. Typical system temperatures were 200—250 K). After combination of the two polarization channels, the new receivers yield an improvement of nearly a factor of 3 in signal-to-noise ratio for line observations at 1mm compared to the old receivers, for the same integration time and spectral resolution.

Where possible, we combined the new extended A configuration data with our previous A and B configuration observations (Tacconi et al. 2006, Greve et al. 2005, Neri et al.

2003) to maximize the sensitivity and UV coverage of the maps, and to ensure that no flux was resolved out at the highest resolution. In the following, the total integration times include the lower resolution configuration observations. For data taken prior to January 2007, the correlator was configured for line and continuum observations to cover simultaneously 580 MHz in each of the 3mm and 1.3mm bands. For observations taken with the new generation receivers (after January 2007) the full available correlator capacity was configured to cover 1 GHz in each polarization. For N2 850.4 the total on source integration was 40 hours (~15 hours in the extended A configuration). For N2 850.2 we only included data from the new receivers, since we did not have usable 1mm data from previous runs, and the new data from the new generation receivers were of such excellent quality. In this case the on-source integration time was 9 hours in the extended A configuration. For the sources in the Hubble Deep Field North, HDF 76 and HDF 242 we combined data from present and past observing runs. The total integration times were 23 hours (12 hours in extended A) and 38 hours (15 hours in extended A), for HDF 76 and HDF 242, respectively.

During the 2005/2006 and 2006/2007seasons we also observed three UV-/optically selected star forming galaxies in the same redshift range as the SMGs. Two of these are so-called 'BX' galaxies, which are selected on the basis of their restframe UV colors and magnitudes ('UV bright'), Q1623 BX453 ($z=2.18$) and Q2343 BX389, ($z=2.17$). These sources were taken from the near-IR long slit spectroscopic sample of Erb et al. (2006a,b); see also Förster Schreiber et al. 2006, Law et al. 2007). The third source, BzK15504, is a so called 's-BzK' star forming galaxy, which is selected based on its K-band magnitude ($K \leq 20$: K-bright) and its UV/optical colors (Kong et al. 2006, Genzel et al. 2006). We observed BX453 and BX389 in the compact D-configuration for 15 and 20 hours, respectively, and BzK15504 for 22 hours in the B and C configurations. We observed the CO (3-2) transition, which was redshifted into the 3mm band for all three sources. We did not detect the CO (3-2) line in any of these three galaxies although we reached very sensitive $2\sigma$ limits to the integrated line flux (at the line centroid and for the velocity width determined from the H$\alpha$ observations) of 0.27, 0.27 and 0.17 Jy km s$^{-1}$ for BX 453, BX389 and BzK15504, respectively. For comparison the integrated CO (3-2) line fluxes of the four SMGs discussed above range between 1 and 2.3 Jy km s$^{-1}$.

We calibrated all datasets using the CLIC program in the IRAM GILDAS package (Guilloteau & Lucas 2000). Passband calibration used one or more bright quasars. Phase and amplitude variations within each track were calibrated out by interleaving reference observations of nearby quasars every 20 minutes. Our absolute positional accuracies are ±0.2 arcsecs or better. The overall flux scale for each epoch was set on MWC 349 (1.05 Jy at 102 GHz (2.9 mm) and 1.74 Jy at 238 GHz (1.3 mm)). After flagging bad and high phase noise data, we created data cubes with natural and uniform weighting using the GILDAS package. The resulting FWHM spatial resolutions were ~1.0" for the 3mm band data (2005/2006), and 0.25-0.5" for the 1.3 mm band.

## 3. Results
### 3.1. The SMGs

Figures 1 through 4 display the millimeter line maps, CO position-velocity maps and spectra in the CO (3-2) line at 3mm, as well as the CO (6-5) or CO (7-6) lines at 1.3mm, for the four SMGs. The CO source brightnesses are consistent with what we have found in Tacconi et al. 2006, and are listed in Table 1 of that paper. As described below, two of the sources, HDF 242 and N2850.4 (Figs 2 and 3) show at least 2 spatially resolved, well-separated components.   The other two sources, HDF 76 and N2850.2 (Figs 1 and 4), are compact, and both have double-peaked profiles indicative of a rotating gaseous disk. Where appropriate we also show overlays of the millimeter line emission on rest frame UV/optical and radio images. The CO (6-5) and (7-6) lines originate from rotational levels >100 K above the ground state. They require substantial temperatures and densities to be populated. Tacconi et al. (2006) have argued on the basis of line shapes and line flux ratios that the interstellar gas in SMGs is probably sufficiently warm ($T_{gas}$>35 K) and dense ($n(H_2)$>$10^{3.5}$ cm$^{-3}$) to reach near thermal equilibrium in these levels. This conclusion is broadly consistent with the CO line flux ratios as a function of J. The effective line brightness temperature ratios CO (6-5)/CO (3-2) and CO (7-6)/CO (3-2) are close to or somewhat below unity (Figures 1, 3, 4), indicating that the turnover to subthermal population occurs near J~5-6, in agreement with Weiss et al. (2005, 2007). In this case, the rotational lines up to this turnover state should provide reasonable tracers of the distribution and kinematics of the overall molecular gas in SMGs while lines from above this state probably select somewhat denser gas (Narayanan et al. 2007). In the following we summarize the generic properties of the distributions and kinematics we derive from our new data. For a more detailed description of individual sources see Tacconi et al. (2006) and references therein.

### 3.1.1 Spatial and kinematic evidence for mergers

We spatially resolve the CO intensity distributions and gas kinematics in all four observed SMGs. Two of our sources (HDF 242, Fig.2) and N2850.4, Fig.3) show well-separated, double or multiple knots on scales of 1"-2.5" (8-20 kpc). In these sources the gas kinematics are complex and unordered (bottom middle panel of Fig.1 and bottom left panel of Fig.2). There is no indication of simple rotation. Velocity differences between source components of 100-200 km s$^{-1}$ are present and can be broadly understood as orbital motion but are smaller than the line widths of the individual components, which reach ~800 km s$^{-1}$ FWHM in the southern component of N2850.4. Comparison of the CO (7-6) emission with the rest-frame UV/optical emission of N2 850.4 (Fig.2 upper left panel) adds to the impression that this system is a very complex system of dense gas, massive star formation and probably also AGN activity spread over a region of <8 kpc in diameter (see Smail et al. 2003, Swinbank et al. 2005 for details). There are strong color gradients in the source. The Optical/UV emission and CO emission have very different morphologies and their peaks are not spatially coincident (even when considering the ±0.5" combined astrometric uncertainties). These differences are most likely the result of high, spatially variable extinction, similar to that seen in dusty z~0 starburst galaxies (e.g. M82: Satyapal et al. 1997, Förster Schreiber et al. 2001) and ULIRGs (e.g. Scoville et al. 2000). Our findings are also in agreement with the predictions of from simulations of massive mergers that aim at matching the properties of SMGs (e.g. Narayanan et al. 2006). N2850.4 is probably a late stage, massive merger.

HDF242 exhibits properties of an earlier stage merger; it shows two distinct galaxy systems separated by ~20 kpc in projection (bottom middle panel of Figure 2). The two components are both prominent in CO and radio continuum emission but only the south-western component is clearly present on the K-band image (Figure 2, left panel). There appears to be a kinematic bridge of CO emission connecting the two components (top and bottom middle panels of Fig.2), demonstrating the physical interaction between the two systems.

Tacconi et al. (2006) discuss a third such widely separated double nucleus system, SMMJ094303+4700 (z=3.34), which has two components (H6 and H7) separated by ~4" (25 kpc).   A fourth SMG with clear indications for being a dynamically interacting pair is SMMJ140103+0252 (Ivison et al 2001, Downes & Solomon 2003, Smail et al. 2005, Nesvadba et al. 2007).  Together with recent Hα observations by Swinbank et al. (2006) that reveal multiple components in another 4 SMGs, these four systems resemble local double-nucleus ULIRGs albeit at higher mass and luminosity (Downes & Solomon 1998, Bryant & Scoville 1999, Scoville et al. 2000, Dasyra et al. 2006).

### 3.1.2 Compact rotating merger remnants

The other two SMGs (HDF76, Fig.1; N2850.2, Fig.4) were the most compact SMGs in the sample studied by Tacconi et al. (2006). They exhibit prominent double-peaked CO profiles (right panels of Figs, 3&4), which are seen in about 40% of the SMGs we have studied so far (Genzel et al. 2003, Greve et al. 2005, Tacconi et al. 2006). Our new observations are beginning to spatially resolve the kinematics even in these compact sources. The CO (6-5) position velocity map of HDF76 (Fig.1, bottom left) shows a significant velocity gradient. The blue and red velocity maxima ($\Delta v=210$ km s$^{-1}$) are separated in the NW-SE direction by 0.18" (1.4 kpc). If this velocity gradient is interpreted as rotation in a compact star forming gas disk, the mass enclosed within 700 pc is $10^{9.6}$ M$_\odot$, for an assumed inclination of 52° (section 4.2). The resulting average mass surface and equivalent hydrogen volume densities are $10^{3.5}$ M$_\odot$pc$^{-2}$ and $10^3$ cm$^{-3}$. The second case is similar but even more extreme. N2850.2 exhibits two emission peaks separated by 400 km s$^{-1}$. The position velocity map (Fig.3 bottom left) shows a marginal spatial separation of 0.05-0.1" (400-800 pc) and the total line emission has an intrinsic size of about 0.2" (1.6 kpc). The corresponding dynamical mass, surface density and volume density are $10^{10}$ M$_\odot$, $10^{4.5}$ M$_\odot$pc$^{-2}$ and $10^{4.2}$ cm$^{-3}$. These surface and volume densities are comparable to those in the dense star forming cores of Galactic molecular clouds, or to the central nuclear gas disk(s) of the extremely compact local ULIRG Arp220 (Downes & Solomon 1998, Sakamoto et al. 1999, Downes & Eckart 2007).

How can a mass of $10^{10}$ M$_\odot$, comparable to that of a 0.1M* galaxy, be compressed to such a small volume? The answer to the analogous question in local ULIRGs is unequivocally that such sources are end stage, major (mass ratio <3:1) mergers where the compression of the gas in the final merger of the two nuclei leads to the most powerful star forming event during the entire merger history (Barnes & Hernquist 1996, Mihos & Hernquist 1994,1996, Downes & Solomon 1998, Bryant & Scoville 1999, Scoville et al. 2000, Genzel et al. 2001, Tacconi et al. 2002, Dasyra et al. 2006). To result in the very compact merger remnants we observe, the merger has to be highly dissipative, with a large gas fraction (see Khochfar & Silk 2006). Bouché et al. (2007) find that the z~2-3 SMGs studied with the PdBI lie in the large v, small R sector of the velocity (v) – size ( R ) plane. This part of the v-r plane is otherwise completely unpopulated, requiring extremely low dark matter halo angular momentum parameters ($\lambda$<0.015, Bouché et al. 2007). Alternatively objects in this location may have started with a larger angular momentum but then lost it as a result of a dissipative merger.

The broad linewidths of these two compact SMGs are also interesting when comparing them to CO bright quasars at similar redshift. It has been known for a few years that high redshift SMGs typically have CO linewidths that are more than a factor of 2 broader than those of QSOs hosts at similar redshift and molecular gas mass (e.g. Greve et al. 2005; Carilli and Wang, 2006). Carilli and Wang (2006) have suggested that this difference is more likely due to an inclination effect, where we are viewing the bright QSOs more face-on than we are the highly obscured SMGs, rather than being due to host mass and/or size differences between the QSO and SMG populations. They cite the Gaussian line profiles and the unobscured view to the AGN as evidence for this. Although the new high resolution SMG CO data presented here cannot unequivocally solve this riddle, the small CO sizes and the double-horned line profiles of HDF76 and N2850.2 do support an inclined disk interpretation for the broad linewidths in these two sources.

In summary, all six well resolved SMGs studied with sub-arcsecond millimeter interferometry so far (the four reported here, as well as SMMJ09431+4700 from Tacconi et al. 2006 and SMMJ140103+0252 from Downes & Solomon 2003) appear to be major mergers in various stages, similar to local ULIRGs.

### 3.1.3 Relationship to compact quiescent galaxies at z~1.4-2.6

Recent HST and ground-based studies show that most of the apparently massive (K≤20-21) passively evolving (='q'uiescent, or 'passively' evolving, or 'red-sequence') galaxies in the same redshift range as the SMGs are remarkably compact (Daddi et al. 2004a,b, 2005, Trujillo et al. 2006, 2007, Zirm et al. 2007, Toft et al. 2007). What is the relationship between these 'q'-BzKs (Daddi et al. 2004a, 2005) and 'q'-DRGs (Zirm et al. 2007, Toft et al. 2007) and the SMG population? Figure 5 shows the comparison between these systems, and with other z~2 star forming galaxies (s-BzK/BX/BM/s-DRG), in the dynamical mass-dynamical surface density plane. To place the q-BzKs/DRGs on this plot, we assumed that $M_*=M_{dyn}$ (zero gas and dark matter fraction). We corrected the published total stellar masses to a Chabrier (2003) IMF (factor 0.6 from a 0.1-100 $M_\odot$ Salpeter IMF, and within 15% of a Kroupa (2001) IMF). For the s-DRGs (where also no dynamical mass measurements are available) we assume that the total masses also include gas, with a gas fraction of $f_{gas}$~0.4, motivated by our earlier SMG results (Greve et al. 2005, Tacconi et al. 2006). For the s-BzK/BX/BM galaxies from the VLT-SINFONI SINS survey we adopt the dynamical mass surface densities of Bouché et al. (2007).

It is evident that SMGs have mass surface densities comparable to those of the quiescent galaxies in the same mass and redshift range. This is especially significant for the two well resolved, compact SMGs discussed above, N2850.2 and HDF76, which are the two topmost black squares in Figure 5. In contrast the star forming BzKs/BX/BM objects are much less dense and lie in the same region as low-z disks, as discussed by Bouché et al. (2007). The star forming DRGs scatter more broadly and include systems similar in density to SMGs and q-BzKs/DRGs. It is thus very plausible to conclude that the star forming SMGs and the quiescent q-BzK/DRGs in the same mass, redshift and surface density range are related by an evolutionary sequence.

## 3.2 UV-/Optically selected galaxies

As mentioned in section 2 we also observed three UV-/optically selected z~2 star forming galaxies. Q2343-BX389 (z=2.1737, $K_s$=20.2) and Q1623-BX453 (z=2.1820, $K_s$=19.8) are from the NIRSPEC survey of Erb et al. (2006a,b). Both are fairly typical representatives of the BX NIRSPEC sample in terms of Hα luminosity and estimated stellar ages (0.4-2.8 Gyrs) but somewhat above the median in stellar mass (5-6x10$^{10}$ M$_\odot$ compared to 3x10$^{10}$ M$_\odot$) and star formation rate (60-100 M$_\odot$ yr$^{-1}$ compared to 30 M$_\odot$ yr$^{-1}$. We refer to Förster Schreiber et al. (2006) for BX 389 and to Law et al. (2007) for BX453 for discussions of the sub-arcsecond Hα integral field spectroscopy of these galaxies. BX389 appears to be an almost edge-on, clumpy rotating disk with a large rotation velocity (275 km s$^{-1}$) and scale length ($R_{1/2}$=6 kpc). Its dynamical mass within $R_{1/2}$ (1x10$^{11}$ M$_\odot$) is amongst the largest in the BX UV-selected sample observed as part of the VLT-SINFONI SINS survey (Förster Schreiber et al. 2006, Cresci et al. in prep., Bouché et al. 2007). In contrast BX453 appears to be a dispersion dominated ($\sigma$=90 km s$^{-1}$), fairly compact ($R_{1/2}$=1.7 kpc) source of modest mass (1.7x10$^{10}$ M$_\odot$ within $R_{1/2}$; Law et al. 2007). The optically selected source BzK15504 (z=2.383, $K_s$=19.2) is in all respects fairly typical of the bright end (K<20) of the s-BzK population (Daddi et al. 2004b, Kong et al. 2006, see Supplementary Material in Genzel et al. 2006). Genzel et al. (2006) found with adaptive optics assisted integral field spectroscopy that this source has a large, globally unstable, clumpy gas disk with rotation velocity of 230 km s$^{-1}$ and scale length 4.5 kpc. Within that radius its dynamical mass is 6.5x10$^{10}$ M$_\odot$.

Based on their extinction corrected H$\alpha$ surface brightness the star formation surface densities in all three galaxies are ~1 M$_\odot$yr$^{-1}$ kpc$^{-2}$ and their total star formation rates are ~60-200 M$_\odot$ yr$^{-1}$. Application of the Schmidt-Kennicutt relation (Kennicutt 1998) implies gas surface densities of ~10$^2$-10$^3$ M$_\odot$pc$^{-2}$ and gas masses of a few 10$^{10}$ M$_\odot$. For these values, assuming the same CO luminosity to gas mass conversion factors as for the SMGs implies that all three galaxies should have been easily detectable with the PdBI.

We did not detect the CO (3-2) line in any of these three galaxies and determined 2$\sigma$ limits to the integrated line flux (at the line centroid and for the velocity width determined from the H$\alpha$ observations) of 0.27, 0.27 and 0.17 Jy km s$^{-1}$ for BX 453, BX389 and BzK15504, respectively. These limits are 3-20 times lower than the integrated line fluxes of the almost two dozen SMGs observed so far (0.8-3.5 Jy km

s$^{-1}$, Smail et al. in prep.). For the same CO to gas mass conversion factors these limits would imply remarkably low gas fractions (less than a few percent), which is difficult to understand given their large star formation rates (Erb et al. 2006 a,b). Alternatively, the conversion factor may be different from that in the SMGs. We explore this possibility further in the next section and in Appendix A. Recently Daddi et al (2008) have successfully detected CO emission from two z~1.5 BzK galaxies, implying molecular gas masses of several x$10^{10}$- $10^{11}$ M$_\odot$ in these sources. These detections make our non-detections all the more puzzling, and it is not clear whether different intrinsic gas masses, star formation modes, CO-H$_2$ conversion factors, or some combination of these can account for the differences. Observations of a larger sample of UV/optically selected high redshift star forming galaxies are urgently needed to make further progress in determining the physical properties of these systems.

## 4. IMF and the CO-H$_2$ conversion factor at z~2

One of the issues we wish to address with our observations is to see if we can constrain the form of the IMF in high-z star forming galaxies. In what follows we discuss a first attempt to carry out such a test on the basis of a global comparison of stellar, gas and dynamical masses in those nine z~2-3 SMGs, BX/s-BzKs and LBGs (see Table 2), where measurements or significant limits for all three of these input quantities are presently available.

### 4.1 Dynamical Masses

For this purpose we derived dynamical masses within the half light radius from virial estimates (see Neri et al. 2003, Baker et al. 2004, Erb et al. 2006a,b) or rotation curve models (Förster Schreiber et al. 2006, Genzel et al. 2006), including corrections for inclination. For BX389 we applied the rotation curve modeling of Förster Schreiber et al. (2006), and for BzK15504 the modeling of Genzel et al. (2006). For the sources lacking published velocity gradients (BX453, cB58, 'Cosmic Eye') we applied the usual isotropic virial estimator (e.g. Spitzer 1987),

$$M_{dyn,virial}(R \leq R_{1/2}) = \frac{5\sigma^2 R_{1/2}}{G} \qquad (1),$$

where $\sigma = \Delta v_{FWHM}/2.35$ is the 1-dimensional velocity dispersion ($\Delta v_{FWHM}$ is the FWHM integrated line width) and $R_{1/2}$ the half light radius. For all four SMGs we applied an average of the isotropic estimator above and the 'global rotating disk estimator' introduced by Neri et al. (2003), corrected for $<\sin(i)> = \pi/4 = 1/1.273^3$ in velocity and $<\sin^2(i)> = 2/3$ in mass, resulting in

$$M_{dyn,disk}(R \leq R_{1/2}) \approx 6 \times 10^4 \Delta v_{FWHM}^2 (km/s) R_{1/2}(kpc) \qquad (M_\odot) \qquad (2).$$

We then extrapolated to the total dynamical mass (of the star forming regions of the galaxy) by multiplying by a factor of 2 for the above estimates, or as appropriate for the more complete (and accurate) rotation curve modeling.

## 4.2 Gas Masses

We estimated gas masses from the observed CO (3-2) line luminosities, with a suitable CO line luminosity to $H_2$ (+He) mass conversion factor, and assuming a brightness temperature ratio of CO (3-2) to (1-0) of unity, motivated by the findings of Weiss et al. (2005, 2007). Clearly, observations of the CO J=1-0 transition with the VLA or GBT would be very useful to check this assumption (e.g. Hainline et al. 2006; Riechers et al.

2006). As discussed in more detail in Appendix A and Downes et al. (1993), the relationship between integrated CO line flux $F_{CO}$ (Jy km s$^{-1}$) and total $H_2$ (+He) cold gas mass can be expressed as

$$M_{gas}/M_\odot = 1.75 \times 10^9 \left(\frac{\alpha}{\alpha_G}\right)\left(\frac{F_{CO}}{\text{Jy km/s}}\right)\left(\frac{T_{CO(3-2)}/T_{CO(1-0)}}{1}\right)(1+z)^{-3}\left(\frac{\lambda_{obs}}{\text{mm}}\right)^2\left(\frac{D_L}{\text{Gpc}}\right)^2 \quad (3),$$

where $T_{CO(3-2)}$ and $T_{CO(1-0)}$ are the effective brightness temperatures of the CO (3-2) and (1-0) lines, respectively, $\lambda_{obs}$ is the wavelength of the observed (3-2) line, $D_L$ is the luminosity distance of the galaxy and $\alpha/\alpha_G$ is a conversion factor in units of that of the Milky Way disk. Appendix A discusses the current state of knowledge of this conversion factor in different Galactic and extragalactic environments in the local Universe. There is fairly strong evidence from various measurements as well as theoretical considerations that the conversion factor in starburst galaxies, galactic nuclei and ultra-luminous mergers with large gas surface densities is significantly smaller than the one in the Milky Way disk (Downes and Solomon 1998, Scoville, Yun & Bryant 1997). On the other hand, there is also evidence that the conversion factor may increase with decreasing metallicity, at least for the more diffuse molecular gas of nearby galaxies (e.g. Israel 2000, 2005). In our parameter search discussed below we adopted several plausible combinations of conversion factors that are suggested from these empirical z=0 data.

## 4.3 Stellar Masses

We have modeled the observed spectral energy distribution of the four SMGs discussed above, as well as the z~3 Lyman Break galaxies cB58 (Baker et al. 2004) and the

---

[3] here we have used $<\sin\theta> = (\iint \sin\theta \, d\theta \, d\varphi)/(\iint d\theta \, d\varphi) = \int_0^{\pi/2} \sin\theta \, d\theta$

'Cosmic Eye' (Coppin et al. 2007; Smail et al. 2007) and the z~2 star forming galaxies BX453, BX389 and BzK15504, with the stellar population synthesis technique described in Förster Schreiber et al. (2004). The data used includes optical and near-infrared photometry from the literature for all sources (see captions to Figures 6-8), and mid-infrared fluxes for a subset. Briefly, the modeling procedure determines the stellar evolutionary model with a synthetic spectrum that best reproduces the observed photometry. We used the recent models by Charlot & Bruzual (2007, in prep.), which feature, among other things, an improved treatment of the thermally-pulsing asymptotic giant branch (AGB) phase compared to the previous Bruzual & Charlot (2003) models (see Bruzual 2007). The new models are now in good agreement with those of Maraston et al. (2006). However, in reality we do not find great differences between the stellar masses obtained with the new Charlot & Bruzual models (2007, in prep) and those obtained with the 2003 models. We adopted the reddening prescription of Calzetti et al. (2000) and ran models for 0.2, 0.5 and 1 times solar metallicity. The free parameters were the stellar age, the interstellar extinction, and the flux scaling factor that determines the stellar mass, star formation rate, and luminosities. The flux scaling factor was fit using all photometric points. We ran models for a range of star formation histories, including an instantaneous burst ("single stellar population"), models with exponentially declining star formation rates with 1/e timescales ranging from of 10 Myr to 1 Gyr, and models with constant star formation rates. We used a Chabrier (2003) stellar initial mass function, which is equivalent to a Kroupa (2001) IMF to within about 15%. The best-fit properties and 68% confidence intervals for each source and set of fixed parameters were derived from 200 Monte Carlo simulations, perturbing the input photometry within the $1\sigma$ measurement uncertainties assuming they are Gaussian.

Figures 6 – 8 show the best fitting models, all of which assume a constant star formation history, and Table 2 lists the best fitting parameters. For a given IMF, star formation history, and set of stellar tracks the stellar masses are constrained to between ±20 and ±50%, similar to other published work in this field (Shapley et al. 2001, 2005, Förster Schreiber et al. 2004, Borys et al. 2005, Erb et al. 2006a,b, Papovich et al. 2007). The stellar masses could be more uncertain if a large fraction of the stars were hidden behind a large amount of extinction such that the UV-IR SEDs are not sensitive to this component. While we cannot exclude such a scenario in the very dusty SMGs, we proceed in the following with the assumption that the SED fitting traces most of the stellar mass. Stellar masses range between 0.5 and 25 $\times 10^{10}$ $M_\odot$. Continuous star formation models or models with decay times ≥100 Myrs are favored by the data for the SMGs and BzK/BX galaxies. The visual extinctions $A_V$ range between 1 and 2 magnitudes. Since the broad-band photometry is dominated by the light from stars of all masses present in the galaxy, the star formation rates inferred from the stellar evolutionary modeling tend to represent past-averaged values in comparison to estimates of "instantaneous" star formation rates from Hα or the infrared luminosities, which are dominated by the most massive hot stars. Motivated by the (super)-solar metallicity found from the $R_{23}$ emission line analysis for SMMJ140103+0252 (Tecza et al. 2004; cf. Nesvadba et al. 2007) we adopted solar metallicity tracks for the four SMGs. For the three z~2 UV/optically selected star forming galaxies the emission line ratios of Erb et al. (2006a,b,c) and Förster Schreiber et al. (2006, and in prep) suggest slightly sub-solar metallicities (0.5-0.9 solar). We thus took stellar tracks of 0.5 and 1 times solar metallicities and averaged the results. For the z~3 LBGs we used averages of the results for 0.2 and 0.5 solar metallicity tracks, as the emission line ratios suggest 0.3-0.5 times solar metallicity (Pettini et al. 2000, Smail et al. 2007, Coppin et al. 2007).

## 4.4 Constraints on IMF, CO conversion factors and gas fractions

Table 2 summarizes the modeling inputs and results for all 9 galaxies, including our best estimates of uncertainties. These latter are generally determined by various systematic effects and parameter correlations, especially between stellar masses and star formation histories/ages. We then calculated the global differences between dynamical masses and the sum of gas and stellar masses as a function of IMF and CO to gas conversion factors. We allowed for a contribution of dark matter within the ≤10 kpc radius of the star forming disk/merger, $f_{dark}$=0.1-0.2, motivated by the findings in the local Universe (Gerhard et al. 2001, Kassin et al. 2006). We computed in the usual manner the sample $\chi^2$,

$$\chi^2 = \sum_{i=1}^{9} \left( \frac{M_{dyn}(1-f_{dark}) - M_{gas}(\alpha) - M_*(f_*)}{\sqrt{\delta M_{dyn}^2 + \delta M_{gas}^2 + \delta M_*^2}} \right)_i^2 \qquad (4),$$

where the gas mass is a function of the adopted CO-gas conversion factor α, and the stellar mass is a function of $f_*$=$M_*$/$M_*$(Chabrier), the ratio of the stellar mass to that of a Chabrier IMF with the same rest-frame optical luminosity. The denominator contains the estimate of the uncertainty of each of these mass values. The sum of stellar, gas and dark mass has to be equal to the dynamical mass. This means that for the appropriate IMF and α values, the numerator must be close to zero and the value of $\chi^2$ in equation (3) close to the number of galaxies (=9). Hence the best values for α and $f_*$ can in principle deduced by minimizing $\chi^2$ as a function of these two parameters.

Figure 9 shows the resulting $\chi^2$ distribution as a function of $f_*$ and for different plausible values of $\alpha$ and $f_{dark}$. While the results clearly still have large statistical and systematic uncertainties, this new method appears quite promising and delivers interesting initial constraints.

- For all plausible choices of the CO to gas conversion factors the $\chi^2$ minimum lies between $f_*\sim0.75$ and 1.05, in good agreement with a 'universal' Kroupa or Chabrier IMF. Formally the lowest minimum is achieved for $f_*=0.75$, perhaps indicating an IMF that is somewhat shallower or more top-heavy than a Chabrier/Kroupa IMF. Including a range of $\delta\chi^2=\pm2$ from these minima we find that $f_*$ must lie between 0.5 and 1.3. A 0.1-100 Salpeter IMF ($f_*\sim1.7$) as well as extreme top-heavy IMFs applicable over the total star formation histories of these galaxies appear to be excluded by our data.
- A Galactic conversion factor for all 9 sources ($\alpha_{SMG}=\alpha_{UV/optical}=4.8$) is strongly disfavored as it leads to an overestimate of the dynamical masses for all stellar mass functions and thus to very poor $\chi^2$ values.
- A ULIRG conversion factor ($\alpha_{SMG}=\alpha_{UV/optical}\sim1$) for both SMGs and UV-/optically selected galaxies is possible in terms of our $\chi^2$ analysis. However, this choice would imply gas fractions of only a few percent on average in the BX/BzK/LBG galaxies we have observed. In that case we would have to be observing all these sources at the time when they are about to run out of gas. This appears very improbable and is also inconsistent with the finding of Daddi et al. (2007) that the dispersion of the luminosities/star formation rates of s-BzKs at a given stellar mass is fairly small, implying that the star formation process in these galaxies is steady and has a high duty cycle.

- The lowest $\chi^2$ values are obtained for a combination of CO to gas conversion factors similar to those favored in the local Universe, for instance $\alpha_{SMG}$=1 M$_\odot$ (K km/s pc$^2$)$^{-1}$ ( corresponds to $\alpha/\alpha_G$=0.2), similar to what is found for local ULIRGs, and $\alpha_{UV/optical}$=4.8 ($\alpha/\alpha_G$~1).

- With this selection of conversion factors the gas fractions of UV-/optically selected and submillimeter galaxies appear to be quite similar, which would be the intuitive guess based on their gas and star formation surface densities in the context of a Schmidt-Kennicutt star formation recipe (Bouché et al. 2007). The inferred average gas fractions are ~30% for the SMGs, ~50% for the LBGs and ≤20% for the sBzK/BX galaxies we have observed.

- A Perhaps surprising result emerging from the fitting of the four SMGs is the relatively old stellar ages (0.64-2.5 Gyrs) deduced in the framework of the continuous star formation model. Borys et al. (2005) have likewise found ~2 Gyr ages for their sample of 11 HDFN galaxies (HDF 76 and HDF242 were common between the two studies). These values are comparable to the average ages of the s-BzK population (Daddi et al. 2007) and the K≤20 BX population (Shapley et al. 2004, Erb et al. 2006a,b). The specific values deduced, however, depend on the assumptions about the star formation histories. For exponential decay 'burst' models ages are lower (a few hundred Myrs), but such models in turn lead to larger discrepancies between the current star formation rates deduced from these models and the instantaneous star formation rates inferred from the submillimeter flux densities.

Our conclusions necessarily need to be considered with caution, because of the small sample of sources, the large uncertainties and the assumptions that went into our analysis. For instance we assumed that there is a single IMF for both the strongly bursting (SMG) as well as the more quiescently star forming stellar components and that both have approximately the same extinction. However, if bursting and quiescent star formation modes have different IMFs (as assumed by Baugh et al. 2005), and/or if the obscuration of the stars powering the far-infrared luminosity is sufficiently large, a top-heavy IMF could still be compatible with the burst component. Another area of concern is the accuracy of the stellar masses obtained from SED fitting. While the statistical errors can be reasonably well estimated with the methods we have applied, the inherent uncertainties in the input stellar models (notwithstanding recent improvements, see above), star formation histories and dust extinction of different stellar components are much harder to quantify and certainly would tend to increase our error bars. A final unknown is the fraction of dark matter in the central few kpc of the galaxies we studied. With these caveats in mind the current evidence supports a near universal (=Galactic) IMF in the z~2 star forming galaxies we have observed, in excellent agreement with the discussion of Renzini (2005). Our measurements also favor $^{12}$CO to gas mass conversion factors that are dependent on ISM gas surface density $\Sigma_{gas}$ (and thus on gas pressure P~$(\Sigma_{gas})^2$) and on metallicity. Obviously an extension of our 'dynamical' method to a larger sample of galaxies, and including more systems with good rotation curves to constrain baryonic and dark matter contributions would be highly desirable.

## 5. Discussion and Conclusions

We have presented in this paper <0.5" FWHM millimeter interferometry observations of four z~2 SMGs and strict limits to the CO fluxes of three UV-/optically selected z~2 galaxies. We have spatially resolved the kinematic structure in the former set of compact, powerful star forming galaxies. We find strong dynamical evidence for major dissipative merging. Velocity fields are complex and chaotic and there are interacting double systems. Several SMGs have low angular momentum gas disks. Our data thus add compelling empirical evidence in favor of previous arguments (e.g. Smail et al. 2002, 2005, Greve et al. 2005, Swinbank et al. 2006, Tacconi et al. 2006) that SMGs are short-duration maximum starburst events in the evolution of a major, gas rich (>30% gas fraction) merger of massive galaxies. But how long do these SMG events last?

We can constrain the duration and duty cycle of the SMG phase in two ways. First, we can make a quantitative estimate of the duration of the starburst phase of these objects from the ratio of the cosmic volume density of SMGs and that of the quiescent population in the same mass and redshift ranges, as the latter are plausibly the descendants of the former (3.1). Table 1 shows a comparison of the cosmic co-moving volume densities of the different samples taken from the recent literature. The quiescent objects are highly clustered (Kong et al. 2006), resulting in large cosmic variance. The volume densities listed in Table 1 are derived from observations in a number of fields, however, so that the amplitude of the cosmic variance should be reduced, which is reflected in the error bars quoted in Table 1. Comparing the density of K≤20 objects to the SMGs, we estimate the SMG 'duty cycle' to be $\eta_{SMG}$~7%. This is justified for two reasons. First a selection of passive galaxies to this limit picks galaxies with $M_* \geq 10^{11}$ $M_\odot$, similar to the stellar mass of $S_{850\mu m} \geq 5$ mJy SMGs (Smail et al. 2004, Borys et al. 2005, Table 2). Second the clustering correlation length of $S_{850\mu m} \geq 5$ mJy SMGs is (6.9+/-2.1)$h^{-1}$ Mpc (Blain et al. 2004), in good agreement with the clustering correlation length of K≤20 q-BzKs (Kong et al. 2006, Daddi et al. in prep). If one adds in addition to SMGs the sample of optically-faint radio galaxies in the same z~2-3 redshift range, which plausibly are gas rich major mergers of somewhat hotter dust temperature but otherwise of star formation rate and mass comparable to those of the $S_{850\mu m} \geq 5$ mJy SMGs (Chapman et al. 2004, Smail et al.

2004), $\eta_{SMG}$ would go up by a factor of two or so. On the other hand if one compares to q-DRGs to K≤21.7, $\eta_{SMG}$ would be lower by a factor of 2-4, depending what fraction of these systems still is in the same mass range as the SMGs. We conclude that a plausible estimate of the duty cycle of M~$10^{11}$ $M_\odot$ major merger induced starbursts is $\eta_{SMG}$~10%, with an uncertainty of a factor of two in both directions. Over the ~1 -1.5 Gyr duration of the cosmic epoch considered here this translates into a duration of the SMG phase of 100-150 Myrs, or 50-300 Myrs taking into account the uncertainties.

A second, independent estimate of the SMG duty cycle comes from the comparison of stellar ages and gas exhaustion time scales. From the gas masses and instantaneous star formation rates (from submm fluxes) in Table 2 we compute the current gas exhaustion time scales, $t_{exhaust}$~$2M_{gas}$/ SFR (last column in Table 2). We then infer the duty cycle $\eta_*$ of the current star formation activity from $\eta_*=2t_{exhaust}/t_*$, where $t_*$ is the age of the stellar population inferred from our SED modeling (second to last column in Table 2), and the factor 2 takes into account in an average sense the 'previous' star formation history. With the CO-$H_2$ conversion factors discussed in the last section we find $\eta_*$(SMG)~0.1. Allowing for additional gas accretion after the current SMG event would increase the gas reservoir, and thus also $\eta_*$. The estimates of the SMG duty cycle based on volume densities and gas exhaustion time scales thus are in good agreement with each other, and also with earlier estimates (Tecza et al. 2004, Smail et al. 2004, Chapman et al. 2005, Greve et al. 2005, Swinbank et al. 2006). The same considerations indicate much larger duty cycles for the UV-/optically selected z~2 galaxies in our sample ($\eta_*$(sBzK/BX)≤0.75 and $\eta_*$(LBG)~1). This is in agreement with Daddi et al. (2007), who find a relatively small scatter of the star formation rate to stellar mass ratio of z~1.5-2.5 s-BzKs, implying that star formation in s-BzKs is fairly steady (see also Noeske et al. 2007 for a similar conclusion at z~1). The short duration of the SMG phase likely results from the combined requirements of triggering such a 'maximum starburst' (Tacconi et al 2006) and terminating it owing to negative feedback from supernova and AGN activity (Chapman et al. 2005). The ~100 Myr total duration is also broadly consistent with the duration of peri-center passage compressions and the final merger phase of a dissipative major merger at z~2 (Mihos & Hernquist 1994, 1996, Barnes & Hernquist 1996, Hopkins et al. 2006, Chakrabarti et al. 2007). Keeping in mind the large uncertainties in model parameters, extinction corrections, IMF etc. we propose that a significant fraction (~50%) of the stellar mass of these ~$M_*$ galaxies formed and assembled in the most active phases of a single gas rich, major merger event, while the rest formed over a longer period of time (> 1Gyr), both during as well as before the merger. In that case one would naturally expect a significant fraction of those SMGs that are not in the submm bright state but have not yet lost most of their gas to be present in the s-BzK/DRG/BX samples, in agreement with Reddy et al. (2005). They may be similar to other s-BzK/DRG/BX galaxies in terms of mass and star formation rate but recognizable as 'mergers' in terms of disturbed kinematics and gas distributions.

The observations presented above add new information on the formation and evolution of massive spheroids. If the properties of the small SMG sample we have studied so far are typical for the SMG population as a whole, we have found the smoking gun for an (or possibly the most) important creation process of spheroids at high redshift: a major and highly dissipative, 'wet' merger of very gas rich galaxies, leading to rapid and efficient star formation and compact merger remnant formation. This conclusion is not surprising. A merger origin of the SMG phenomenon has been favored for some time (e.g. Smail et al. 1998, 2004, Chapman et al. 2004, Khochfar & Silk 2006, Cimatti et al. 2008). Our data, however, provide clear-cut evidence for this assumption (see also Swinbank et al. 2006). Our SMG data also support other emerging evidence from high resolution optical observations that many or most of the massive spheroids at z~1.4-2.5 are much more compact and dense than z~0 spheroids (Daddi et al. 2005, Zirm et al. 2007, Toft et al. 2007, Trujillo et al. 2007, 2008). Cimatti et al. (2008) conclude from the first spectroscopic data for z~1.5 q-BzKs that their stellar ages are about 1 Gyr, consistent with a formation redshift of z~2-3. What becomes of the merger remnants between z~2 and z~0? To connect the z~2 location of the SMGs/q-DRGs/q-BzKs to the z~0 spheroid track, the further evolutionary path has to proceed from high density to low density (Figure 7). Dry major mergers would appear to be a natural choice (van Dokkum 2005, Bell et al. 2006, Naab, Khochfar & Burkert 2006, Boylan-Kolchin, Ma & Quataert 2006). The dry merger track (red arrow in the right panel of Figure 5, Nipoti et al. 2003) connects the locus of the z~2 remnants to the most massive cluster ellipticals at z~0 at a mass of $M_* \geq 10^{12}\ M_\odot$. The z=0 comoving volume density of dark halos with the appropriate mass for such clusters (~$10^{14}\ M_\odot$) is ~$2.5 \times 10^{-5}\ h_{0.7}^3$ Mpc$^{-3}$ (Mo & White 2002), sufficiently large to be consistent with the hypothesis that every SMG at z~2-3 becomes a massive cluster elliptical.

However in the most simple version of dry major mergers, the dense central regions of the original compact high-z remnants would have to still be observable at z~0, in contradiction with the inward extrapolation of a Hernquist (1990) model for a very massive z~0 remnant that could be the end result of this process (dotted curves in the right inset of Fig.5). In addition, Genel et al. (in prep.) have carried out an analysis of the Millennium dark matter simulation (Springel et al. 2005), which indicates that a halo of mass ~$10^{12}$-$10^{13}\ M_\odot$, appropriate for a SMG, experiences on average only ~1 major merger between z~2.5 and z=0, and only ~8% of these halos experience 3 or more major mergers in this redshift interval. This major merger rate is not sufficient to account for the required factor ~5-10 mass increase from the z~2 SMGs/q-BzKs/q-DRGs onto the z=0 spheroid track. Finally, Burkert, Naab & Johansson (in prep.) conclude from the structure/dynamics of massive and boxy, z~0 spheroids in the SAURON sample that their properties cannot be accounted for solely by major dry mergers but more probably require a sequence of major *and minor* mergers. Further detailed simulations are needed to settle the question whether a combination of dry major mergers, minor mergers as well as smooth accretion lead to the observed properties of very massive local ellipticals (Cox et al. 2006, Boylan-Kolchin, Ma & Quataert 2006, Naab et al. 2007).

Another possibility is that the compactness of the z~2 SMG/q-BzK/q-DRG population is misleading in terms of their overall stellar mass distribution. In reality there may be a bright central starburst region dominating the surface brightness distribution, which is surrounded by a much larger halo of older stars dominating the overall mass distribution (c.f. Daddi et al. 2005). As discussed by Tacconi et al. (2002) in the context of compact z~0 ULIRG merger remnants, the half light radius of such a configuration would be expected to significantly increase over a few Gyrs, as initially the most massive OB stars (dominating the dust/CO emission) and then the bright TP-AGB stars (dominating the q-BzK/q-DRG phase) in the central starburst fade over time. This effect obviously would lessen the discrepancy between the half light densities of the z~2 and z~0 spheroids.

*Acknowledgements.* We thank the staff of the IRAM Observatory for their support of this program. We are grateful to Thorsten Naab, Andi Burkert, and Amiel Sternberg for valuable discussions. We also thank the referee for constructive comments that have helped to clarify and improve the paper. IRS acknowledges support from the Royal Society, and AMS acknowledges support from STFC.

# Appendix A. The CO to $H_2$ conversion factor

More than 30 years of molecular line observations in giant molecular clouds of the Milky Way have established that the integrated line flux of $^{12}$CO millimeter rotational lines can be used to infer gas masses, despite the facts that the CO molecule only makes up a small fraction of the entire gas mass and that the lower rotational lines (1-0, 2-1, 3-2) are almost always very optically thick (Solomon et al. 1987, and references therein). This result stems from the fact that CO emission in the Milky Way and nearby normal galaxies comes from moderately dense (volume averaged densities $<n(H_2)> \sim 200$ cm$^{-3}$), virialized (i.e. self-gravitating) giant molecular clouds (GMCs). In this regime it can be shown (Dickman et al. 1986, Solomon et al. 1987, Solomon & Barrett 1991) that the ratio of $H_2$ column density to CO flux $I_{CO}$ ($I_{CO}=\int_{line} T_R(v)dv$), or gas mass (including a 36% mass correction for helium) to CO luminosity $L'_{CO}$ ($L'_{CO}=\int_{source} \int_{line} T_R(v) \, dv \, dA$) can be expressed as

$$N(H_2)/I(CO) = X = c_1 \left( \frac{<n(H_2)>}{200 \text{ cm}^{-2}} \right)^{1/2} \left( \frac{T_R}{6 \text{ K}} \right)^{-1} \quad , \quad [\text{cm}^{-2}/(\text{K km s}^{-1})] \quad (A1)$$

and

$$M_{gas}/L'_{CO} = \alpha = c_2 \left( \frac{<n(H_2)>}{200 \text{ cm}^{-2}} \right)^{1/2} \left( \frac{T_R}{6 \text{ K}} \right)^{-1} \quad , \quad [M_\odot /(\text{K km s}^{-1}\text{pc}^2)] \quad (A2).$$

Here $T_R$ is the equivalent Rayleigh Jeans brightness temperature of the (optically thick) CO line and $c_1$ and $c_2$ are appropriate numerical constants. In the 2.6mm CO (1-0) transition the typical gas temperature of Galactic GMCs is ~8-10 K. For spherical clouds supported by isotropic random motions (e.g. turbulence) in virial equilibrium with their own gravity, the resulting inferred theoretical conversion factors are $X \sim 2 \times 10^{20}$ [cm$^{-2}$ (K km s$^{-1}$)$^{-1}$] and $\alpha=4.3$ [M$_\odot$ (K km/s pc$^2$)$^{-1}$]. Several independent empirical techniques based on GeV γ-rays, optical extinction measurements, isotopomeric line ratios and excitation analysis have all shown that this 'virial' technique is appropriate and remarkably robust throughout the Milky Way (Strong & Mattox 1996, Dame et al. 2001, Dickman et al. 1986, Solomon et al. 1987). The 'standard' empirical conversion factor $X_G$ ranges between 2 and 3 x$10^{20}$ (Left blue filled circle in Fig. 10). In the following we adopt $X_G$=2.2x$10^{20}$ and $\alpha_G$=4.8. The validity of this simple proportionality is surprising, as it implies that average densities and temperatures vary little across the Milky Way clouds, or only in a way that the ratio of $n(H_2)^{1/2}/T$ is constant.

Dickman et al. (1986) have extended the virial approach to external galaxies by showing that equations A1 and A2 apply even when one is observing an ensemble of virialized clouds, instead of a single one, as long as again the factor $n(H_2)^{1/2}/T$ is constant throughout the system and the CO line is optically thick. This assumption of an ensemble of individual gas clouds in virial equilibrium with their own gravity breaks down, however, in galactic nuclei and starburst galaxies. Gas motions there are due to a combination of gas and stellar mass components, and the gas is often in a smoother, disk-like configuration. Downes et al. (1993), Solomon et al. (1997) and Downes and Solomon (1998) have shown that in this limit equations A1 and A2 still apply but in a slightly modified form

$$N(H_2)/I(CO) = X = c_3 f_{gas}^{1/2} \left(\frac{H/R}{0.15}\right)^{1/2} \left(\frac{<n(H_2)>}{200 \text{ cm}^{-2}}\right)^{1/2} \left(\frac{T_R}{6 \text{ K}}\right)^{-1} , \quad [\text{cm}^{-2}/(\text{K km s}^{-1})] \quad (A3)$$

and

$$M_{gas}/L'_{CO} = \alpha = c_4 f_{gas}^{1/2} \left(\frac{H/R}{0.15}\right)^{1/2} \left(\frac{<n(H_2)>}{200 \text{ cm}^{-2}}\right)^{1/2} \left(\frac{T_R}{6 \text{ K}}\right)^{-1} , \quad [M_\odot/(\text{K km s}^{-1}\text{pc}^2)] \quad (A4).$$

Here $f_{gas}$ is the mass fraction of gas in the galaxy, H/R is the ratio of vertical height and radius of the gas layer, and $c_3$ and $c_4$ are numerical constants. For conditions appropriate for luminous or ultraluminous infrared galaxies ((U)LIRGs: $n(H_2)$~$10^3$-$10^4$ cm$^{-3}$ and $T_R$~20-50 K) with H/R=0.15-0.2 and $f_{gas}$~0.1-0.3, the inferred theoretical conversion factors range between α=0.8 and 1.6 (X=3.7-7.3x10$^{19}$: Solomon et al. 1997, Downes & Solomon 1998). Again, various empirical calibrations of the X/α factors in the Galactic Center, nearby AGN and (U)LIRGs, based on dynamical mass measurements, are in good agreement with these theoretical considerations. Figure 10 shows an (incomplete) compilation of a number of these X-factor determinations in the literature. Since in the context of the 'Schmidt-Kennicutt' (Kennicutt 1998) empirical relation star formation surface density scales with gas surface density it is appropriate to plot, as in Figure 10, the derived X-factors as a function of $\Sigma_{gas}$ ($M_\odot pc^{-2}$). These measurements (crossed green squares and the bottom blue filled square for clouds in the Milky Way Center) suggest that in dense environments ($\Sigma_{gas}$~ $10^{2.5}$-$10^4$ $M_\odot pc^{-2}$) the conversion factor is 0.2 to 0.5 times the value in the extended GMC population of the Milky Way disk.

Another important variable is the metallicity of the gas. The above discussion relates to near- or super-solar metallicity environments in moderately massive z~0 galaxies. Using calibrations based mainly on optically thin far-infrared/submillimeter dust emission and isotopomeric line measurements several groups have found that the global conversion factor in nearby galaxies appears to be a strong function of metallicity (Arimoto et al. 1996, Israel 2000, 2005, Boselli et al. 2002). The average of these results gives the following relationship between X-factor and oxygen abundance

$$\log(X) = 29.2 - (12 + \{O/H\}) \qquad (A5).$$

This relationship is plotted in Figure 10 as filled black triangles for oxygen abundances of 9.2 (bottom) and 8.2 (top); the arrow connects these extreme values.

Rosolowsky et al. (2003) and Leroy et al. (2006) have derived X-factors from high resolution, virial measurements toward individual dense GMCs in nearby galaxies over the same range in metallicity as discussed above. These measurements indicate only a weak metallicity dependence of the X-factor (filled red squares in Figure 10). This is in contrast to and probably superseding earlier interferometric measurements by Wilson (1995) that did show a trend similar to equation A5.

This situation is confusing. However, the contradiction may be qualitatively understood and resolved in terms of the physics of photodissociation regions (e.g. Israel 2000). In star forming galaxies the optically thick CO lines (1-0, 2-1, 3-2 etc.) come from the surface layers of GMCs that are exposed to and heated by the local far-UV radiation field (Wolfire, Hollenbach & Tielens 1993). Far-UV radiation also dissociates CO molecules but in solar metallicity environments dust shielding and CO self-shielding are efficient enough that only a moderately thin (~10%) surface layer is dissociated where carbon is then in the form of CII and CI. At lower metallicities, dust and CO self shielding is correspondingly less efficient. For instance, for $Z \sim 0.25\, Z_\odot$ UV radiation dissociates 6-10 times deeper in terms of overall hydrogen column density than at solar metallicity, while the much more effective $H_2$ self shielding leaves the $H_2$ column basically unaffected (Maloney & Black 1988, Maloney & Wolfire 1997). In dense enough clouds, the remaining CO column is typically still sufficiently large to make the lower rotation transitions of $^{12}$CO optically thick, resulting in similar conversion factors as in higher metallicity environments. For the more diffuse molecular gas the situation may be drastically different, however, leading to a situation that the effective filling factor of CO emitting gas is much less than that of $H_2$ (Israel 2000). These considerations may explain the difference between the interferometric measurements of Rosolowsky et al. (2003) and Leroy et al. (2006) and the global measurements of Arimoto, Israel, Boselli and others. They are also consistent with the far-infrared observations of the [CII] 158μm line by Madden et al. (1997) who infer the presence of molecular gas without CO emission in the Z~0.25 $Z_\odot$ galaxy IC10.

How are the values in Figure 10 to be extrapolated to high redshift and what conversion factors are likely applicable to SMGs or UV-/optically selected z~2-3 star forming galaxies? Optical emission line ratios (F([NII])/F(Hα), $R_{23}$=(F(OIII)+F(OII))/F(Hβ)) suggest that the high mass end (K<20) of the UV-/optically selected z~2 star forming galaxies ($M_*$~$10^{11}$ $M_\odot$), including BX389, BX453 and BzK15504 in Table 2, have ionized gas phase metallicities close to but somewhat below solar ((12 + {O/H})~8.4-8.6, Erb et al. 2006c, Shapley et al. 2004, Förster Schreiber et al., in prep.). The metallicities of the z~3 LBGs are lower (Pettini et al. 2001). For cB58 Pettini et al. (2000) find (12 +{O/H}~8.1). BzK15504 (Genzel et al. 2006) and BX389 (Förster Schreiber et al. 2006) are rotating disks with large, self-gravitating gas complexes so that equations A1 and A2 are probably appropriate. For the more dispersion dominated sources BX453, and perhaps cB58 (Baker et al. 2004) and the 'Cosmic Eye' (Coppin et al. 2007) as well, both sets of equations A1/2 and A3/4 may be appropriate (see discussion in Coppin et al. 2007). Typical gas volume and surface densities of Galactic molecular clouds are $10^{2.5}$ cm$^{-3}$ and $10^{2.5}$ $M_\odot$ pc$^{-2}$, and gas temperatures may be 20-30 K. From these considerations we conclude that for BX389, BX453 and BzK15504 a conversion factor of α/$α_G$~1 is a plausible guess. Given their lower metallicities the LBGs may have α/$α_G$~1-3 (see Baker et al. 2004). The SMGs are scaled-up, and more gas-rich versions of local ULIRGs (Tacconi et al. 2006, Greve et al. 2005), with average molecular gas densities of >$10^3$ cm$^{-3}$, gas surface densities >$10^3$-$10^4$ $M_\odot$ pc$^{-2}$ and gas temperatures of 30-40 K. In the case of SMG14011+0252 J1 at z=2.56, Tecza et al. (2004) find (12+{O/H})=8.92±0.07 from the $R_{23}$ metallicity estimator, significantly super-solar. If these values are representative for the SMGs as a whole, as may be suggested by the [NII]/H$\alpha$ line ratios found by Swinbank et al. (2004), an appropriate conversion factor is $\alpha/\alpha_G$~0.2-0.5, similar to the values for local starbursts and ULIRGs in Figure 10.

# References

Adelberger, K.L., Steidel, C.C., Shapley, A.E., Hunt, M.P., Erb, D.K., Reddy, N.A. & Pettini, M. 2004, ApJ, 607, 226

Arimoto, N., Sofue, Y. & Tsujimoto, T. 1996, PASJ, 48, 275

Baker, A., Tacconi, L.J., Genzel, R., Lehnert, M.D. & Lutz, D. 2004, ApJ, 604, 125

Baugh, C. M., Lacey, C. G., Frenk, C. S., Granato, G. L., Silva, L., Bressan, A., Benson, A. J. & Cole, S. 2005, MNRAS, 356, 1151

Barnes, J.E. & Hernquist, L. 1996, ApJ, 471, 115

Bell, E. et al. 2006, ApJ, 640, 241

Biggs, A.D & Ivison, R.J. 2007, MNRAS, in press (arXiv:0712.3047v1)

Blain, A.W., Smail, I., Ivison, R.J., Kneib, J.-P. & Frayer, D.T. 2002, Phys.Rep., 369, 111

Blain, A.W., Chapman, S.C., Smail, I.R. & Ivison, R.J. 2004, ApJ, 611, 725

Borys, C., Smail, I., Chapman, S. C., Blain, A. W., Alexander, D. M.& Ivison, R. J. 2005, ApJ, 635, 853

Boselli, A., Lequeux, J. & Gavazzi, G. 2002, Ap&SS, 281, 127

Bouché, N. et al. 2007, ApJ, 671, 303

Boylan-Kolchin, M., Ma, C.-P. & Quataert, E. 2006, MNRAS, 369, 1081

Bruzual, G. & Charlot, S. 2003, MNRAS, 344, 1000

Bruzual, G. 2007, arXiv:0702.091v1

Bryant, P.M. & Scoville, N.Z. 1999, AJ, 117, 2632

Calzetti, D., Armus, L., Bohlin, R.C., Kinney, A.L., Koorneef, J. & Storchi-Bergmann, T. 2000, ApJ, 533, 682

Carilli, C.L. & Wang, R. 2006, AJ, 131, 2763

Chabrier, G. 2003, PASP, 115, 763

Chakrabarti, S., Cox, T.J., Hernquist, L., Hopkins, P., Robertson, B. & di Matteo, T. 2007, ApJ, 658, 850

Chapman, S.C., Blain, A.W., Smail, I.R. & Ivison, R.J. 2005, ApJ, 622, 772

Chapman, S.C., Smail, I.R., Windhorst, R., Muxlow, T. & Ivison, R.J. 2004, ApJ, 611, 732

Chapman, S.C., Blain, A.W., Ivison, R.J. & Smail, I.R. 2003, Nature, 422, 695

Cimatti, A. et al. 2008, A&A, in press (arXiv:0801.1184v1)

Coppin, K.E.K. et al. 2006, MNRAS, 372, 1621

Coppin, K.E.K. et al. 2007, ApJ, 665, 936

Cox, T.J., Dutta, S.N., di Matteo, T., Hernquist, L., Hopkins, P.F., Robertson, B. & Springel, V. 2006, ApJ, 650, 791

Daddi, E. et al. 2004a, ApJ, 600, L127

Daddi, E., Cimatti, A., Renzini, A., Fontana, A., Mignoli, M., Pozetti, L., Tozzi, P. & Zamorani, G. 2004b, ApJ, 617, 746

Daddi, E. et al. 2005, ApJ, 626, 680

Daddi, E. et al. 2007, ApJ, 670, 156

Daddi. E. et al. 2008, ApJ, 673, L71

Dame, T. M., Hartmann, D. & Thaddeus, P. 2001, ApJ, 547, 792

Dasyra, K.M., Tacconi, L.J., Davies, R.I., Genzel, R., Lutz, D., Naab, T., Burkert, A., Veilleux, S. & Sanders, D.B. 2006, ApJ, 638, 745

Davé, R. 2007, MNRAS, in press (arXiv:0710.0381v2)

Davies, R., Tacconi, L.J. & Genzel, R. 2004, ApJ, 602, 148

Dickman, R.L., Snell, R.L. & Schloerb, P.F. 1986, ApJ, 309, 326

Downes, D., Solomon, P.M. & Radford, S.J.E. 1993, ApJ, 414, L13

Downes, D. & Solomon, P. M. 1998, ApJ, 507, 615

Downes, D. & Solomon, P.M. 2003, ApJ, 582, 37

Downes, D. & Eckart, A. 2007, A&A, 468, L57

Ellingson, E., Yee, H. K. C., Bechtold, J., Elston, R. & Carlberg, R. G. 1996, J.Roy.Astr.Soc.Can., 90, 313

Erb, D.K., Steidel, C.C., Shapley, A.E., Pettini, M., Reddy, N.A. & Adelberger, K.L. 2006a, ApJ, 646, 107

Erb, D.K., Steidel, C.C., Shapley, A.E., Pettini, M., Reddy, N.A. & Adelberger, K.L. 2006b, ApJ, 647, 128

Erb, D.K., Shapley, A.E., Pettini, M., Steidel, C.C., Reddy, N.A. & Adelberger, K.L. 2006c, ApJ, 644, 813

Förster Schreiber, N. M., Genzel, R., Lutz, D., Kunze, D. & Sternberg, A. 2001, ApJ, 552, 544

Förster Schreiber, N.M. et al. 2004, ApJ, 616, 40

Förster Schreiber, N.M. et al. 2006, ApJ, 645, 1062

Franx, M., et al. 2003, ApJ, 587, L79

Genzel, R., Tacconi, L.J., Rigopoulou, D., Lutz, D. & Tecza, M. 2001, ApJ, 563, 527

Genzel, R., Baker, A. J., Tacconi, L. J., Lutz, D., Cox, P., Guilloteau, S. & Omont, A. 2003, ApJ, 584, 633

Genzel, R. et al. 2006, Nature, 442, 786

Gerhard, O., Kronawitter, A., Saglia, R.B., & Bender, R. 2001, AJ, 121, 1936

Grazian, A. et al. 2007, A&A, 465, 393

Greve, T.R., Bertoldi, F., Smail, I., Neri, R., Blain, A.W., Ivison, R.J., Chapman, S.C., Genzel, R., Omont, A., Cox, P. Tacconi, L.J. & Kneib, J.-P. 2005, MNRAS, 359, 1165

Guilloteau, S. & Lucas, R. 2000, in Imaging at Radio through Submillimeter Wavelengths, ed. J. G. Mangum & S. J. E. Radford (San Francisco: ASP), 299

Guilloteau, S. et al. 1992, A&A, 262, 624

Hainline, L.J., Blain, A.W., Greve, T.R., Chapman, S.C., Smail, I. & Ivison, R.J. 2006, ApJ, 650, 614

Harayama, Y., Eisenhauer, F. & Martins, F. 2007, ApJ, in press (arXiv:0710.2882v2)

Hernquist, L. 1990, ApJ, 356, 359

Hinz, J.L. & Rieke, G.H. 2006, ApJ, 646, 872

Hopkins, P.F., Hernquist, L., Cox, T.J., di Matteo, T., Robertson, B. & Springel, V. 2006, ApJS, 163, 1

Hughes, D. et al. 1998, Nature, 394, 241

Israel, F.P. 2000, in Molecular Hydrogen in Space, Eds. F.Combes & G.Pineau de Foréts, Cambridge : Cambridge Univ.Press, 326

Israel, F.P. 2005, A&A, 438, 855

Ivison, R.J, Smail, I. Frayer, D.T., Kneib, J.-P. & Blain, A.W. 2001, ApJ, 561, L45

Ivison, R.J., Greve, T.R., Smail, I., Dunlop, J.S., Roche, N.D., Scott, S.E., Page, M.J., Stevens, J.A., Almaini, O., Blain, A.W., Willott, C.J., Fox, M.J, Gilbank, D.G., Serjeant, S., & Hughes, D.H. 2002, MNRAS, 337, 1

Kassin, S.A., de Jong, R.S. & Weiner, B. 2006, ApJ, 643, 804

Kennicutt, R.C. Jr. 1998, ApJ, 498, 541

Khochfar, S. & Silk, J. 2006, ApJ, 648, L21

Kong, X. et al. 2006, ApJ, 638, 72

Kroupa, P. 2001, MNRAS, 322, 231

Lacey, C. G., Baugh, C. M., Frenk, C. S., Silva, L., Granato, G. L. & Bressan, A. 2007, MNRAS, in press (arXiv:0704.1562)

Law, D.R., Steidel, C.C., Erb, D. K., Larkin, J.E., Pettini, M., Shapley, A.E. & Wright, S.A., 2007, ApJ, 669, 929

Leroy, A., Bolatto, A., Walter, F. & Blitz, L. 2006, ApJ, 643, 825

Madden, S.C., Poglitsch, A., Geis, N., Stacey, G.J. & Townes, C.H 1997, ApJ, 483, 200

Maloney, P. & Black, J.H. 1988, ApJ, 325, 389

Maloney, P.R. & Wolfire, M.G. 1997, in IAU Symposium 170, 299

Maraston, C., Daddi, E., Renzini, A. et al. 2006, ApJ, 625, 85

Mihos, J.C. & Hernquist, L. 1994, ApJ, 431, L9

Mihos, J.C. & Hernquist, L. 1996, ApJ, 464, 641

Mo, H. J. & White, S. D. M. 2002, MNRAS, 336, 112

Naab, T., Khochfar, S. & Burkert, A. 2006, ApJ, 636, L81

Naab, T., Johansson, P.H., Ostriker, J.P. & Efstathiou, G. 2007, ApJ, 657, 710

Nagashima, M., Lacey, C. G., Okamoto, T., Baugh, C. M.. Frenk, C. S. & Cole, S. 2005, MNRAS, 363, L61

Narayanan, D., Cox, T.J., Robertson, B., Dave, R., di Matteo, T., Hernquist, L., Hopkins, P.F., Kulesa, C. & Walker, C.K. 2006, ApJ, 642, L107

Narayanan, D., Cox, T.J., Kelly, B., Dave, R., Hernquist, L., Di Matteo, T., Hopkins, P., Kulesa, C., Robertson, B. & Walker, C.K. 2007, ApJ, submitted (arXiv:0710.0384)

Nayakshin, S. & Sunyaev, R. 2005, MNRAS, 364, L23

Neri, R., Genzel, R., Ivison, R.J., Bertoldi, F., Blain, A.W., Chapman, S.C., Cox, P., Greve, T.R., Omont, A. & Frayer, D.T. 2003, ApJ, 597, L113

Nesvadba, N.P.H. et al. 2007, ApJ, 657, 725

Nipoti, C., Londrillo, P. & Ciotti, L. 2003, MNRAS, 342, 501

Noeske, K.G. et al. 2007, ApJ, 660, L43

Oka, T., Hasegawa, T., Hayashi, M., Handa, T., & Sakamoto, S. 1998, ApJ, 493, 730

Papovich, C. et al. 2007, ApJ, 668, 45

Paumard, T. et al. 2006, ApJ, 643, 1011

Pettini, M., Steidel, C.C., Adelberger, K.L., Dickinson, M. & Giavalisco, M. 2000, ApJ, 528, 96

Pettini, M., Shapley, A.E., Steidel, C.C., Cuby, J.-G., Dickinson, M., Moorwood, A.F.M., Adelberger, K.L. & Giavalisco, M. 2001, ApJ, 554, 981

Pope, A., Borys, C., Scott, D., Conselice, C., Dickinson, M. & Mobasher, B. 2005, MNRAS, 358, 149

Reddy, N. A., Erb, D. K.,Steidel, C. C., Shapley, A. E., Adelberger, K. L. & Pettini, M. 2005, ApJ, 633, 748

Renzini, A. 2005, in 'The Initial Mass Function 50 Years Later', eds. E. Corbelli, F. Palle & H. Zinnecker, Astrophysics and Space Science Library, 327, 221

Riechers, D.A. et al. 2006, AJ, 650, 604

Rosolowsky, E., Engargiola, G., Plambeck, R.L. & Blitz, L. 2003, ApJ, 599, 258

Sanders, D.& Mirabel, I.F. 1996, ARAA, 34, 749

Sakamoto, K., Scoville, N.Z., Yun, M.S., Crosas, M., Genzel, R. & Tacconi, L.J. 1999, ApJ, 514, 68

Satyapal, S., Watson, D.M., Pipher, J. L., Forrest, W. J., Greenhouse, M. A., Smith, H. A., Fischer, J. & Woodward, C. E. 1997, ApJ, 483, 148

Scoville, N. Z., Yun, M. S. & Bryant, P. M. 1997, ApJ, 484, 702

Scoville, N.Z., Evans, A.S., Thompson, R., Rieke, M.J., Hines, D.C., Low, F.J., Dinshaw, N., Surace, J.A. & Armus, L. 2000, AJ, 119, 991

Shapley, A.E., Steidel, C.C, Adelberger, K.L, Dickinson, M., Giavalisco, M. & Pettini, M. 2001, ApJ, 562, 95

Shapley, A.E., Erb, D.K., Pettini, M., Steidel, C.C., & Adelberger, K.L. 2004, ApJ, 612, 108

Shapley, A.E., Steidel, C.C., Erb, D.K., Reddy, N.A., Adelberger, K.L., Pettini, M., Barmby, P. & Huang, J. 2005, ApJ, 626, 698

Shen, S., Mo, H.J, White, S.D.M., Blanton, M.R., Kauffmann, G., Voges, W., Brinkmann, J & Csabai, I 2003, MNRAS, 343, 978

Shier, L.M., Rieke, M.J. & Rieke, G.H. 1994, ApJ, 433, L9

Smail, I. Ivison, R.J., Blain, A.W. & Kneib, J.-P. 1998, ApJ, 507, L21

Smail, I., Ivison, R. J., Blain, A. W. & Kneib, J.-P. 2002, MNRAS, 331, 495

Smail, I., Chapman, S.C., Ivison, R.J., Blain, A.W., Takata, T., Heckman, T.M., Dunlop, J.S. & Sekiguchi, K. 2003, MNRAS, 342, 1185

Smail, I., Chapman, S.C., Blain, A.W. & Ivison, R.J. 2004, ApJ, 616, 71

Smail, I. & Smith, G.P. & Ivison, R.J. 2005, ApJ, 631, 121

Smail, I. et al. 2007, ApJ, 654, L33

Solomon, P.M., Rivolo, A.R., Barrett, J. & Yahil, A. 1987, ApJ, 319, 730

Solomon, P.M. & Barrett, J.W. 1991, in IAU Symposium 146, eds. F.Combes & F.Casoli, Dordrecht: Kluwer, 235

Solomon, P.M., Downes, D., Radford, S.J.E. & Barrett, J.W. 1997, ApJ, 478, 144

Spitzer, L.Jr, 1987, in Dynamics and Evolution of Globular Clusters, Princeton: Princeton University Press.

Springel, V. et al. 2005, Nature, 435, 629

Steidel, C. C., Giavalisco, M., Pettini, M., Dickinson, M. & Adelberger, K. L. 1996, ApJ, 462, L17

Steidel, C.C., Shapley, A.E., Pettini, M., Adelberger, K.L., Erb, D.K., Reddy, N.A. & Hunt, M.P. 2004, ApJ, 604, 534

Stolte, A., Grebel, E.K., Brandner, W. & Figer, D.F. 2002, A&A, 394, 459

Strong, A.W. & Mattox, J.R. 1996, A&A, 308, L21

Swinbank, A.M., Smail, I., Chapman, S.C., Blain, A.W., Ivison, R.J. & Keel, W.C. 2004, ApJ, 617, 64

Swinbank, A.M., Smail, I., Bower, R.G., Borys, C., Chapman, S.C., Blain, A.W., Ivison, R.J., Ramsay Howat, S., Keel, W.C. & Bunker, A.J. 2005, MNRAS, 359, 401

Swinbank, A.M., Chapman, S.C., Smail, I., Lindner, C., Borys, C., Blain, A.W., Ivison, R.J. & Lewis, G.F. 2006, MNRAS, 371, 465

Tacconi, L. J., Genzel, R., Lutz, D., Rigopoulou, D., Baker, A. J., Iserlohe, C. & Tecza, M. 2002, ApJ, 580, 73

Tacconi, L.J. et al. 2006, ApJ, 640, 228

Tecza, M., Baker, A.J., Davies, R.I., Genzel, R., Lehnert, M.D., Eisenhauer, F., Lutz, D., Nesvadba, N., Seitz, S., Tacconi, L.J., Thatte, N.A., Abuter, R. & Bender, R. 2004, ApJ, 605, L109

Toft, S. et al. 2007, ApJ, 671, 285

Trujillo, I. et al. 2006 MNRAS, 373, L36

Trujillo, I., Conselice, C., Bundy, K., Cooper, M.C., Eisenhardt, P. & Ellis, R.S. 2007, MNRAS, 382, 109

van Dokkum, P. 2005, AJ, 130, 2647

van Dokkum, P. 2007, ApJ, in press (arXiv:0710.0875v3)

Weiss, A., Neininger, N., Hüttemeister, S. & Klein, U. 2001, A&A, 365, 571

Weiss, A., Downes, D., Walter, F. & Henkel, C. 2005, A&A, 438, 533

Weiss, A., Downes, D., Walter, F. & Henkel, C. 2007, in 'from z-machines to ALMA: Submillimeter spectroscopy of galaxies', eds. A.J. Baker, J.Glenn, A.I. Harris, J.G. Mangum & M.S.Yun, Asp Conf.series, 375, 25

Wolfire, M., Hollenbach, D.J. & Tielens, A.G.G.M. 1993, ApJ, 402, 195

Wild, W. et al. 1992, A&A, 265, 447

Wilson, C.D. 1995, ApJ, 448, L97

Yee, H. K. C., Ellingson, E., Bechtold, J., Carlberg, R. G. & Cuillandre, J.-C. 1996, AJ, 111, 1783

Zirm, A.W. et al. 2007, ApJ, 656, 66

# Figure Captions

SMMJ123549+6215 (HDF76) z=2.20

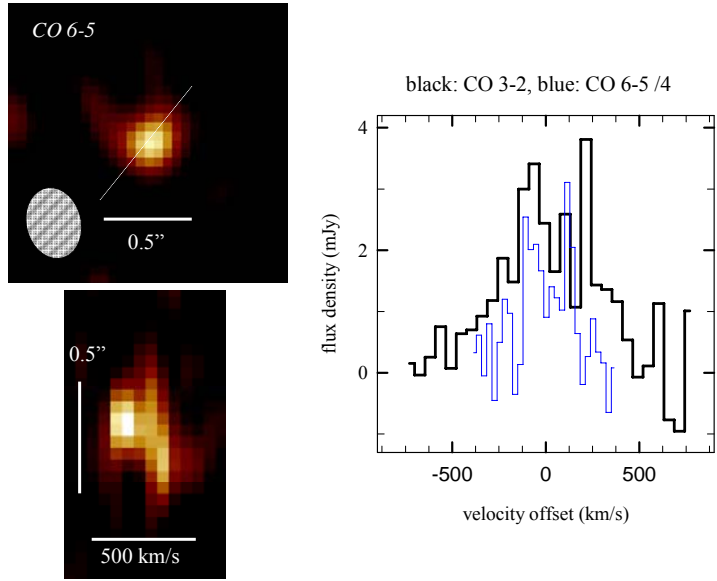

Figure 1. Integrated CO (6-5) line map (top left panel) and position velocity diagram (bottom left panel) along position angle -42$^0$ (dotted line in top panel) in the source HDF76 at z=2.2. The angular resolution is 0.42"x0.29" at a PA=-168° FWHM (hatched ellipse). The position velocity diagram was smoothed with a 100mas/110 km s$^{-1}$ Gaussian kernel. The integrated CO (3-2) spectrum (black, from Tacconi et al. 2006) and CO (6-5) spectrum (blue, scaled by 0.25 to be on the same Rayleigh-Jeans brightness temperature scale) is shown in the right panel.

# SMMJ123707+6214 (HDF 242) z=2.49

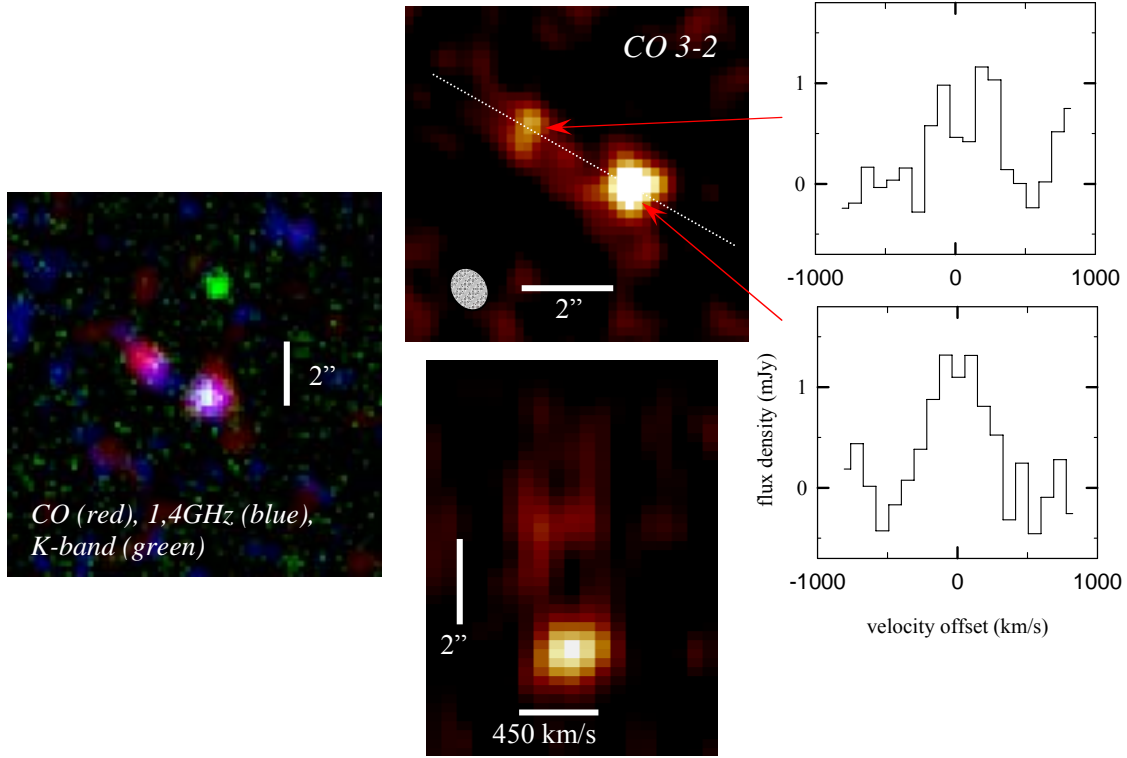

Figure 2. Integrated CO (3-2) line map (top central panel, angular resolution of 1"x0.86" at PA=37° FWHM: hatched ellipse) and position velocity diagram (bottom central panel) at position angle $60^0$ (dotted line) of the source HDF242 at z=2.49. The position velocity diagram was smoothed with a 0.4"/185 km s$^{-1}$ Gaussian kernel. The right panels show CO (3-2) profiles of the two most prominent CO concentrations. The left panel shows a three color composite, with CO (3-2) in red, 1.4GHz VLA continuum in blue (Biggs & Ivison 2007) and K-band in green (Smail et al. 2004). The south-western CO peak is prominent also in the rest-frame optical and radio, while the secondary north-eastern CO concentration is only strongly seen in the radio continuum map. The relative astrometry of these images is accurate to ±0.5".

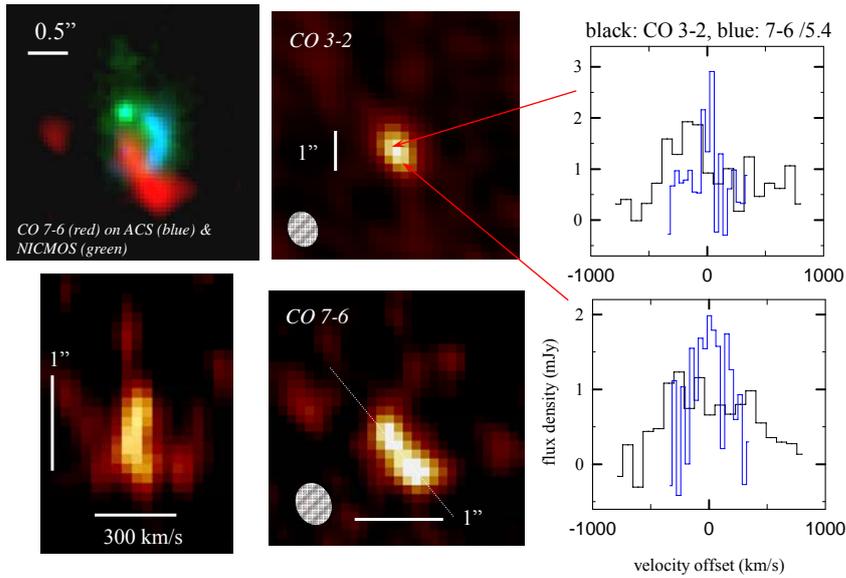

Figure 3. CO (3-2) (top central, 0.9" x 0.8" at PA 27° FWHM, hatched ellipse ) and CO (7-6) (bottom central, 0.52"x 0.4" at PA=27° FWHM, hatched ellipse) integrated line maps and CO (7-6) position velocity diagram (bottom left) along the dotted line in the bottom central inset, in the source N2850.4 at z=2.39. The position velocity diagram was smoothed with a 0.2"/74 km s$^{-1}$ Gaussian kernel. The two right panels show CO (3-2) profiles (black) and CO (7-6) profiles (blue) toward the south-western CO (7-6) peak (bottom) and north-western CO (7-6) peak (top), showing a similar kinematic structure as in the CO (7-6) position velocity diagram. The CO (7-6) flux densities are scaled by 0.19 to bring CO (7-6) and (3-2) profiles on the same Rayleigh-Jeans brightness temperature scale. The top left inset shows a three color composite of the CO (7-6) emission in red, the ACS image in blue and the NICMOS image in green (Swinbank et al. 2005). The relative astrometry of these images is accurate to ±0.5".

SMMJ16358+4105 (N2850.2) z=2.45

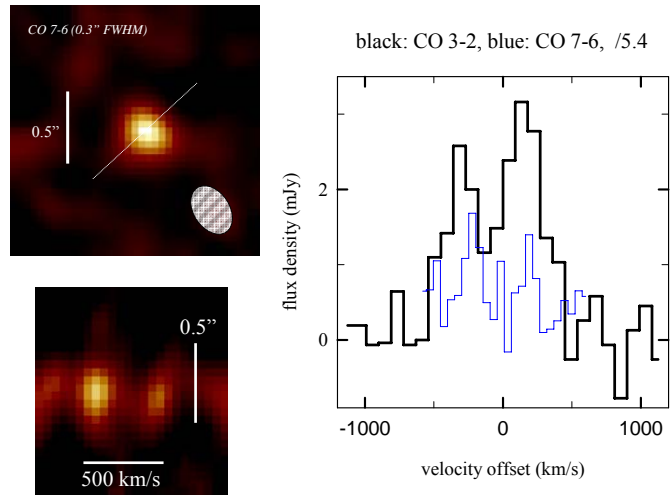

Figure 4. Integrated CO (7-6) line map (top left panel, 0.36"x0.25" FWHM at PA=39°, hatched ellipse) and position velocity cut (bottom left) in the direction marked by a dotted line in the integrated flux map, of the source N2850.2 at z=2.45. The position velocity diagram was smoothed with a 0.1"/110 km s$^{-1}$ Gaussian kernel. The right panel shows the integrated CO (3-2) (black: Tacconi et al. 2006) and 7-6 profiles (blue, scaled by 0.19 to put both lines on the same Rayleigh-Jeans brightness temperature).

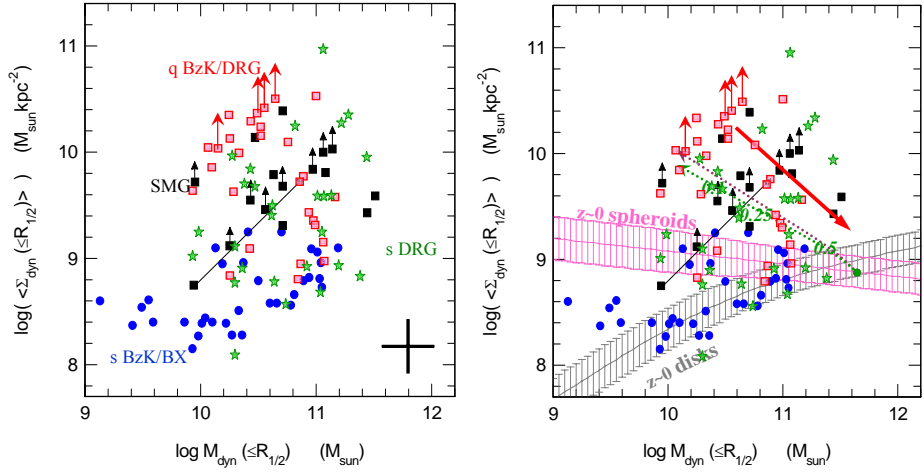

Figure 5. Comparison of the star forming SMGs (z=1-3.5) and various rest-frame UV-/optically identified galaxy populations, in the dynamical mass-surface density plane. The horizontal axis is the dynamical mass within the half light radius, $M_{dyn}(R \leq R_{1/2})$. The vertical axis is the mass surface density within the half light radius, $\Sigma_{dyn} = M_{dyn}(R \leq R_{1/2})/(\pi R_{1/2}^2)$. In the plots blue filled circles mark the location of the star forming BzK/BX/BM galaxies (z=1.4-2.5) from the VLT-SINS survey (Bouché et al. 2007). Black squares denote the location of the SMGs from our PdBI survey (Bouché et al. 2007, this work). Arrows mark those SMGs where the present PdBI data only establish upper size limits (Tacconi et al. 2006, Smail et al. 2007), or where, in the case of SMMJ140103+0252, the source is probably face-on, implying a large inclination correction to the intrinsic circular velocity (long diagonal arrow, Nesvadba et al. 2007). Green stars mark star forming ('s')-DRGs and red squares mark quiescent q-BzKs/DRGs from Daddi et al. 2005, Trujillo et al. 2006, Zirm et al. 2007 and Toft et al. 2007. The large black cross in the lower right denotes the typical uncertainty of these points. To bring the q-BzKs/DRGs on this plot of dynamical masses and surface densities, we assumed that $M_*=M_{dyn}$ for these sources but corrected the published total stellar masses to a Kroupa (2001) of Chabrier (2003) IMF by multiplying 0.1-100 $M_\odot$ Salpeter IMF based masses by 0.6. For the s-DRGs we assumed that the dynamical masses also include gas, with a gas fraction of 0.4, motivated by our SMG results (Greve et al. 2005, Tacconi et al. 2006). The hatched pink and grey zones in the right inset mark the location of local spheroids and disks, respectively, from the work of Shen et al. (2003). A red arrow denotes the dry merger line, as discussed in Nipoti, Londrillo and Ciotti (2003). The dotted curves show the enclosed mass (green) and enclosed light (purple) in a Hernquist (1990) profile. Numbers 0.5, 0.25 and 0.1 denote the location of the Hernquist projections within 0.5, 0.25 and 0.1 times the effective radius. The curves shown are for the specific example of a $R_{1/2}=15$ kpc, $M_{tot}=10^{12}$ $M_\odot$ system.

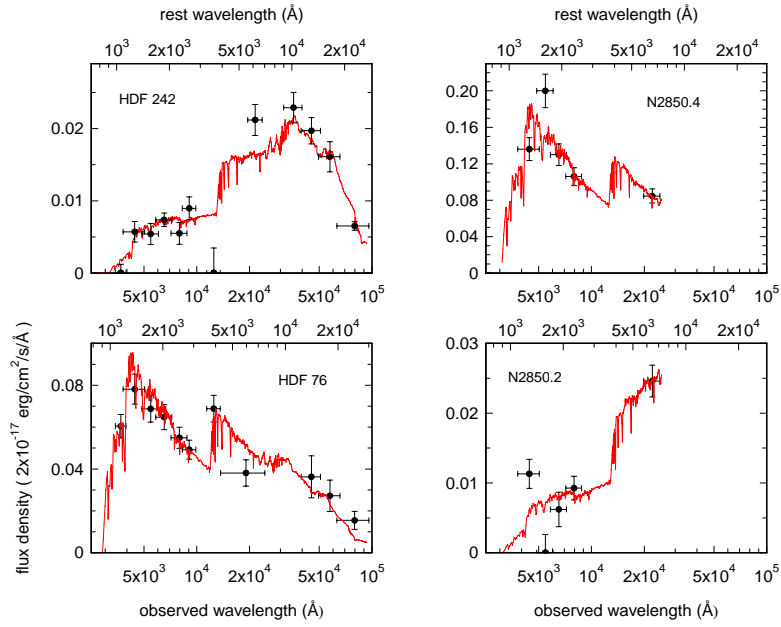

Figure 6. Rest-frame UV- to near-infrared spectral energy distributions of the four SMGs discussed in this paper, along with best fitting constant star formation models and stellar ages, as discussed in the text. The photometric data points are from Borys et al. (2005) for the two HDF sources, and from Smail et al. (2004), Ivison et al. (2002) and Chapman et al. (2005) for the two ELAIS N2 sources. Stellar masses quoted are for a Chabrier (2003) IMF and solar metallicity tracks; they include a correction for foreground extinction by the Milky Way. For the best fitting star formation history parameters and masses see Table 2.

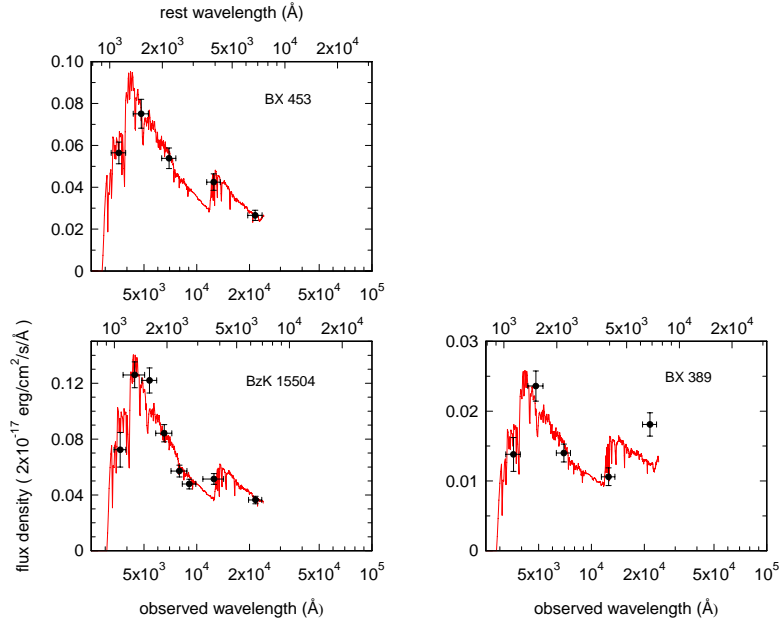

Figure 7. Rest-frame UV- to optical spectral energy distributions of the three z~2 UV-/optically selected sources discussed in this paper, along with best fitting constant star formation models and stellar ages, as discussed in the text. The photometric data points are from Daddi (priv.comm.) for BzK15504 (see also supplementary material in Genzel et al. 2006), and from Erb et al. (2006b) for BX453 and BX389. The outlying K-band data point of BX389 is probably a result of the contribution of Hα. Stellar masses quoted are for a Chabrier (2003) IMF and solar metallicity tracks; they include a correction for foreground extinction by the Milky Way. For the best fitting star formation history parameters and masses see Table 2.

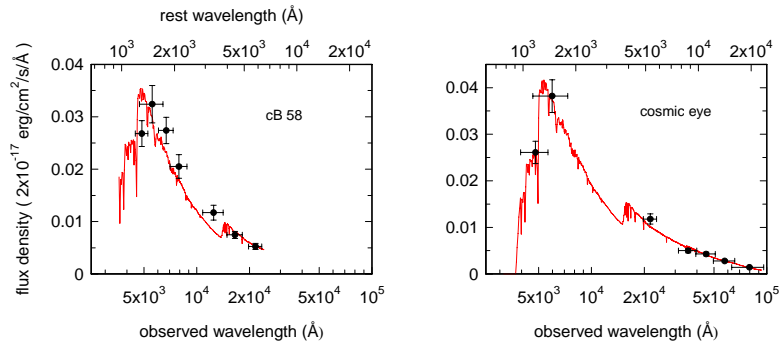

Figure 8. Rest-frame UV- to near-infrared spectral energy distributions of the two lensed Lyman Break galaxies discussed in this paper, along with best fitting constant star formation models and stellar ages, as discussed in the text. The photometric data points are Yee et al. (1996) and Ellingson et al. (1996) for cB58 and from Smail et al. (2007) and Coppin et al. (2007) for the Cosmic Eye. Stellar masses quoted are for a Chabrier (2003) IMF and 0.2 times solar metallicity tracks, and include a correction for foreground extinction by the Milky Way and lensing magnification (factor 31.8 for cB58 and 28 for the Cosmic Eye). For the best fitting star formation history parameters and masses see Table 2.

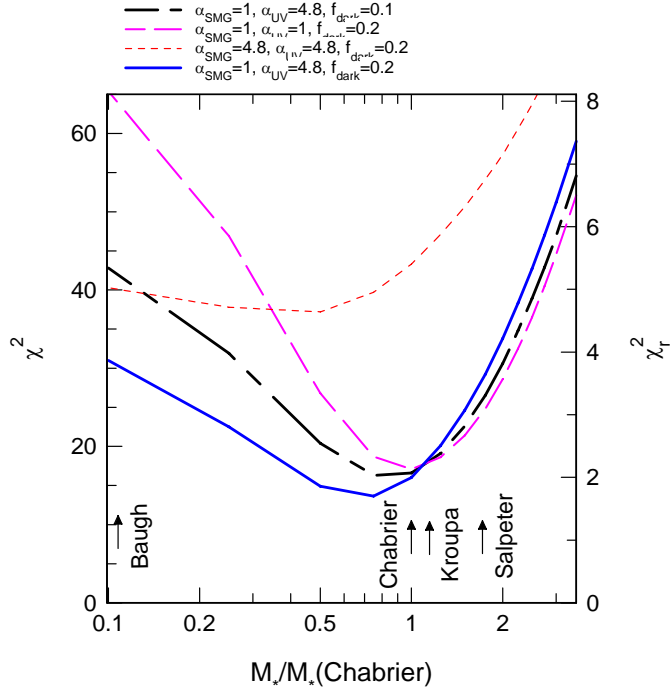

Figure 9. Dependence of $\chi^2=\Sigma_i(M_{dyn}-M_{gas}(\alpha)-M_*(IMF)-M_{dark})/error)_i^2$ as a function of the IMF chosen (horizontal scale, in units of total M/L ratio, relative to the M/L ratio of a Kroupa (2001) IMF, for 9 z~2-3 SMGs, s-BzK/BX and LBGs with good constraints on dynamical masses $M_{dyn}$, gas masses $M_{gas}$ and stellar masses $M_*$. $\chi_r^2 = \chi^2/8$ is the reduced $\chi^2$ value (right vertical scale). Continuous, dashed, dash-dotted and dotted curves are for four different choices of the CO conversion factor $\alpha$ (in units of K km s$^{-1}$ pc$^2$), and for a dark matter contribution of 10 and 20%, as annotated. In the $\chi^2$ computation (equation 3), 2σ limits to gas masses were treated as detections.

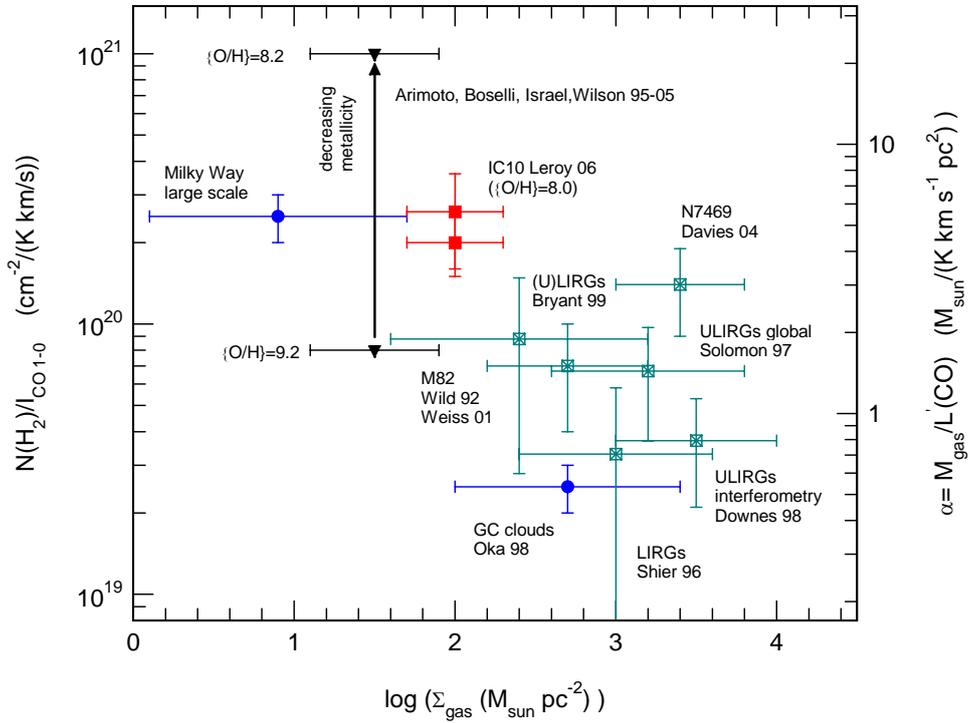

Figure 10. Compilation of the conversion factor X from CO (1-0) flux ($I_{CO}$ (K km s$^{-1}$)) or luminosity ($L'_{CO}$ (K km s$^{-1}$ pc$^2$)) to $H_2$ column density (left vertical scale) and total ($H_2$+HeI) gas mass (right vertical scale), derived in various z~0 Galactic and extragalactic targets. Filled blue circles denote measurements in the disk and center of the Milky Way, based on various virial, extinction and isotopomeric analyses (Solomon et al. 1987, Strong & Mattox 1996, Dame et al. 2001, Oka et al. 1998). Crossed green squares denote measurements in starbursts and (U)LIRGs, mainly based on dynamical constraints (Wild et al. 1992, Davies et al. 2004, Shier et al. 1994, Hinz & Rieke 2006, Weiss et al. 2001, Solomon et al. 1997, Downes & Solomon 1998, Bryant & Scoville 1999). Filled

squares denote conversion factors as a function of decreasing metallicity (vertical arrow) from {O/H}=12 + [O/H]=9.2 (bottom) to 8.2 (top), derived mainly from global (large scale) dust mass measurements in nearby galaxies and dwarfs by several groups (Arimoto et al. 1996, Israel 2000, 2005, Boselli et al. 2002, see also Wilson 1995). In contrast red filled squares mark X-factor measurements toward individual clouds, over the same range in metallicity (Rosolowsky et al. 2003, Leroy et al. 2006).

**Table 1. Comoving volume densities of different z~1-3.4 galaxy samples**

| galaxy sample | $\Phi$ ($h_{70}^3$ Mpc$^{-3}$) |
|---|---|
| submillimeter galaxies (SMGs): $S_{850\mu m} \geq 5$ mJy, z=1-3.4<br>optically faint radio galaxies OFRGs: z=1-3.4 | $1.1 \pm 0.1 \times 10^{-5}$ [a]<br>$\sim 1 \times 10^{-5}$ [a] |
| BX/BM K$\leq$20 z=1.4-2.5<br>star forming K$\leq$20 BzK z=1.4-2.6<br>star forming K$\leq$20 DRG z=2.0-2.6<br>star forming K$\leq$21.7 DRG z=2.0-2.6 | $1.1 \pm 0.3 \times 10^{-4}$ [b,d]<br>$2.2 \pm 0.6 \times 10^{-4}$ [b,c,d]<br>$7.5 \times 10^{-5}$ [b]<br>$12 \pm 3 \times 10^{-4}$ [b,e] |
| quiescent (passive) K$\leq$20 BzK z=1.4-2<br>quiescent (passive) K$\leq$21.7 DRG z=2.0-2.6 | $1.5 \pm 0.4 \times 10^{-4}$ [b,c,d]<br>$6.5 \pm 2 \times 10^{-4}$ [b,e] |

[a] Chapman et al. 2005, [b] Reddy et al. 2005 (GOODS N), [c] Daddi et al. 2004a,b, 2005, Kong et al. 2006 (GOODS S, HUDF, Deep 3a, Daddi-F), [d] Grazian et al. 2007 (MUSIC), [e] Zirm et al. 2007, Toft et al. 2007 (HDF S, MS1054)

**Table 2. Physical properties of z~2-3 star forming galaxies**

| source | Z | $v_c$[a] km/s | $R_{1/2}$[b] kpc | $M_{dyn}(\leq 2R_{1/2})$[c] $10^{10} M_\odot$ | $M_*$[d] $10^{10} M_\odot$ | $M_{gas}(CO)$[e] $10^{10} M_\odot$ | $R_*$[f] $M_\odot$/yr | $R(H\alpha)$[g] $M_\odot$/yr | $R(IR)$[h] $M_\odot$/yr | $t_*$[i] Myr | $t_{exh}$[j] Myr |
|---|---|---|---|---|---|---|---|---|---|---|---|
| HDF 76 | 2.20 | 384 (80) | 0.95 (0.4) | 9.4 (4)[k] | 12 (+4, -3) | 4(0.5)α[k] | 183 (+95, -11) | 255 (20) | 900 (500) | 1000 (600) | 44α |
| HDF242 | 2.49 | 275 (55) | 2.8 (1) | 24 (12) | 12 (+2.5, -4.5) | 2.9(0.5)α | 200 (+400, -65) | 37 (11) | 500 (250) | 1000 (900) | 56α |
| N2850.2 | 2.45 | 510 (100) | 0.8 (0.5) | 14(7) | 25 (7.5) | 5.5(0.6)α | 200 (+70, -44) | 42 (10) | 1100 (500) | 2500 (+0, -900) | 50α |
| N2850.4 | 2.39 | 450 (90) | 2.4 (1) | 34(14) | 23 (+11, -7) | 6.8α (0.88) | 460 (+260, -175) | 190 (30) | 900 (450) | 640 (+1300, -320) | 75α |
| BX Q1623-453 | 2.18 | 140 (73) | 1.7 (0.5) | 3.4(1.2) | 4.0 (2) | <0.7α (2σ) | 130 (70) | 290 (170) | - | 400 (+400, -200) | <63α |
| BX Q2343-389 | 2.17 | 275 (28) | 6.2 (2) | 20(3) | 5 (1.5) | <0.7α (2σ) | 30 (10) | 210 (130) | - | 2750 (1000) | <140α |
| BzK 15504 | 2.38 | 230 (16) | 4.5 (1) | 13(1.5) | 10 (2.1) | <0.5α (2σ) | 120 (+5, -40) | 200 (100) | - | 1280 (+1200, -260) | <34α |
| cB58[l] | 2.73 | 113 (23) | 1.0 (0.2) | 1.3 (0.7) | 0.5 (0.1) | 0.044α (0.0075) | 100 (40) | - | - | 50 (+20,-0) | 4.4α |
| Cosmic Eye[m] | 3.07 | 124 (25) | 1 (0.2) | 1.5(0.9) | 1.5 (0.2) | 0.28α (0.13) | 220 (20) | - | - | 100 (20) | 13α |

[a] inferred peak circular velocity, based on the work of Cresci et al. (in prep), Bouché et al. (2007), Förster Schreiber et al. (2006), Genzel et al. (2006), Tacconi et al. (2006), Law et al. (2007) and this paper. Here and elsewhere, values in parenthesis are 1σ uncertainties.

[b] intrinsic HWHM of Hα/CO emission (and uncertainty), assumed to be identical to scale length of disk. Numbers are based on the work of Cresci et al. (in prep), Bouche et al. (2007), Förster Schreiber et al. (2006), Genzel et al. (2006), Tacconi et al. (2006), Law et al. (2007) and this paper

[c] total dynamical mass (and uncertainty), within $2*R_{1/2}$.

[d] total stellar mass (and uncertainty), from Monte Carlo fitting to the restframe UV-/optical/near-IR spectral energy distribution, assuming a Kroupa (2001)/Chabrier (2003) IMF, stellar tracks from Bruzual and Charlot (2007, in prep.), a Calzetti et al. (2000) extinction law, and a constant or an average of constant/single stellar population star formation history. For the SMGs we adopted solar metallicity, for BzK15504 and the BX galaxies we adopted an average of the results for 0.5 and 1.0 solar metallicity, and for the LBGs we adopted an average for the 0.2 and 0.5 solar metallicity. The photometry is from Borys et al. (2005) for HDF sources, from Ivison et al. (2002), Smail et al. (2004) & Chapman et al. (2005) for the N2850 sources, from Yee et al. (1996) and Ellingson et al. (1996) for cB58 and from Coppin et al. (2007) and Smail et al. (2007) for the 'Cosmic Eye'.

[e] gas ($H_2$+He) mass (and uncertainty), estimated from CO intensity and a conversion factor of $N(H_2)/I_{(CO)}$ =4.6x10$^{19}$ (cm$^{-2}$/ (K km s$^{-1}$) (α= 1 $M_\odot$/(K km s$^{-1}$ pc$^2$)), which is 0.2 times the Milky Way conversion factor. Upper limits are 2σ.

[f] extinction corrected star formation rate (and uncertainty) determined from the stellar fitting above [d], based on a constant star formation history and a Kroupa (2001) IMF, including correction for TP-AGB.

[g] extinction corrected star formation rate based on Hα (and uncertainty), using the Kennicutt (1998) conversion but corrected for a Kroupa (2001) IMF. Note that the Calzetti et al. (2000) extinction prescription implies that the gas is extincted by a factor of 2.3 larger than the stars.

[h] star formation rate (and uncertainty) derived from the 850μm flux density and using the conversion SFR=110* $S_{850\mu m}$(mJy), taken from Pope et al. (2005), for a Chabrier IMF.

[i] stellar age, based on the stellar synthesis fitting discussed above [d]
[j] current gas exhaustion time scale, based on $M_{gas}$/(SFR(FIR), SFR(H$\alpha$), SFR(*))
[k] quantities in parentheses are the 1$\sigma$ uncertainties
[l] corrected for a lensing magnification of 31.8 (Baker et al. 2004)
[m] corrected for a lensing magnification of 28 for the total UV/optical photometry but 8 for the CO emission (Coppin et al. 2007)